\documentclass[twocolumn]{aastex63}
\usepackage{amsmath}

\newcommand{\diff}{d}

\newcommand{\mdot}{\dot{M}_w}
\newcommand{\pdot}{\dot{p}_w}
\newcommand{\Lwind}{\mathcal{L}_w}
\newcommand{\mcloud}{M_{\rm cloud}}
\newcommand{\rcloud}{R_{\rm cloud}}
\newcommand{\REC}{R_{\rm EC}}
\newcommand{\RWeaver}{R_W} 
\newcommand{\PWeaver}{P_W}
\newcommand{\prWeaver}{p_W}
\newcommand{\VWeaver}{V_W}
\newcommand{\REB}{R_{\rm EB}}
\newcommand{\prEB}{p_{{\rm EB}}}

\newcommand{\reff}{\mathcal{R}_{b}}
\newcommand{\rfree}{\mathcal{R}_{\rm f}}
\newcommand{\dotreff}{\dot{\mathcal{R}}_{b}}
\newcommand{\Vbub}{V_{b}}
\newcommand{\Ebub}{E_{b}}
\newcommand{\Pbub}{P_{b}}
\newcommand{\Abub}{A_{b}}
\newcommand{\pr}{p_{r}}
\newcommand{\ms}{M_{*}}
\newcommand{\sfe}{\varepsilon_{*}}
\newcommand{\rhobar}{\bar{\rho}}
\newcommand{\ecool}{\dot{E}_{\rm cool}}

\newcommand{\ratr}{\mathcal{S}}

\newcommand{\lcool}{\ell_{\rm cool}}
\newcommand{\Lmix}{L_{\rm mc}}
\newcommand{\kappaeff}{\kappa_{\rm eff}}
\newcommand{\tcool}{t_{\rm cool}}

\newcommand{\lint}{\mathcal{L}_{\rm int}}
\newcommand{\Msh}{M_{\rm sh}}
\newcommand{\Esh}{E_{\rm sh}}
\newcommand{\Ersh}{E_{\rm r, sh}}
\newcommand{\Etsh}{E_{\rm turb, sh}}

\newcommand{\vw}{{\cal{V}}_w}
\newcommand{\vrel}{v_{\rm rel}}
\newcommand{\rhops}{\rho_{\rm ps}}
\newcommand{\Pps}{P_{\rm ps}}
\newcommand{\vsw}{v_\mathrm{ps}}
\newcommand{\vt}{v_{ t}}
\newcommand{\te}{t_e}
\newcommand{\pc}{\mathrm{\, pc}}
\newcommand{\yr}{\mathrm{\, yr}}
\newcommand{\Myr}{\mathrm{\, Myr}}
\newcommand{\kms}{\mathrm{\, km\, s^{-1}}}
\newcommand{\Kel}{\, \mathrm{K}}
\newcommand{\Msun}{\, \mathrm{M}_\odot}
\newcommand{\ellmin}{\ell_\mathrm{min}}

\received{XXX}
\revised{XXX}
\accepted{XXX}

\submitjournal{ApJ}

\shorttitle{Efficiently Cooled Wind Bubbles: Theory}
\shortauthors{Lancaster et al.}
\begin{document}

\title{Efficiently Cooled Stellar Wind Bubbles in Turbulent Clouds I.\\ Fractal Theory and Application to Star-Forming Clouds}

\correspondingauthor{Lachlan Lancaster}
\email{lachlanl@princeton.edu}

\author[0000-0002-0041-4356]{Lachlan Lancaster}
\affiliation{Department of Astrophysical Sciences, Princeton University, 4 Ivy Lane, Princeton, NJ 08544, USA}

\author[0000-0002-0509-9113]{Eve C. Ostriker}
\affiliation{Department of Astrophysical Sciences, Princeton University, 4 Ivy Lane, Princeton, NJ 08544, USA}

\author[0000-0001-6228-8634]{Jeong-Gyu Kim}
\affiliation{Department of Astrophysical Sciences, Princeton University, 4 Ivy Lane, Princeton, NJ 08544, USA}

\author[0000-0003-2896-3725]{Chang-Goo Kim}
\affiliation{Department of Astrophysical Sciences, Princeton University, 4 Ivy Lane, Princeton, NJ 08544, USA}

\begin{abstract}
Winds from massive stars have velocities of $1000 \kms$ or more, and produce hot, 
high pressure gas when they shock. We develop a theory for the evolution of 
bubbles driven by the collective winds from star clusters early in their 
lifetimes, which involves interaction with the turbulent, dense interstellar 
medium of the surrounding natal molecular cloud. A key feature is the fractal 
nature of the hot bubble's surface.  The large area of this interface with 
surrounding denser gas strongly enhances energy losses from the hot interior, 
enabled by turbulent mixing and subsequent cooling at temperatures 
$T\sim 10^4-10^5 {\rm K}$ where radiation is maximally efficient. Due to the 
extreme cooling, the bubble radius scales differently ($\reff \propto t^{1/2}$) 
from the classical \citet{Weaver77} solution, and has expansion velocity and 
momentum lower by factors of $10-10^2$ at given $\reff$, with pressure lower by 
factors of $10^2 - 10^3$.  Our theory explains the weak X-ray emission and low 
shell expansion velocities of observed sources.  We discuss further implications 
of our theory for observations of the hot bubbles and cooled expanding shells 
created by stellar winds, and for predictions of feedback-regulated star formation 
in a range of environments. In a companion paper, we validate our theory with a 
suite of hydrodynamic simulations.
\end{abstract}


\section{Introduction}
\label{sec:intro}

Star formation is a notoriously inefficient process, with only a 
few percent of the gas mass in galaxies being converted to stellar 
mass over the relevant gravitational timescales \citep[e.g.][]{
KennicuttRev98,KrumholzTan07,Evans09,
Murray11,Vutisalchavakul16,Lee16,Barnes17,Utomo18,Kruijssen19}. 
This inefficiency is thought to be caused and regulated 
by feedback from stars which inject mass, momentum, and energy  
to their surroundings. This injection happens from the scale of 
protostellar outflows in cores 
\citep[e.g.][]{MatznerMcKee00,Bally16,OffnerChaban17}, 
to individual massive stars in their natal clouds 
\citep[e.g.][]{RogersPittard13,Geen15a,Haid18}, 
to star clusters in galactic disks  
\citep[e.g.][]{kok13,CGK_TIGRESS1,hennebelle14,Gatto17},
with many different processes playing important roles. 

On the scale of clouds, feedback is thought to be dominated by 
energy provided to the ISM by massive stars early in their lives 
\citep[e.g.][]{KMBBH19,girichidis20,chevance20a}. This energy can 
take a number of forms: direct radiation pressure 
\citep[e.g.][]{wolfire87,KrumholzMatzner09,Raskutti16}, pressure from warm gas 
heated by photo-dissociating/ionizing radiation leading to ionized 
and neutral outflows \citep[e.g.][]{Whitworth79,franco94,Matzner02,JGK18,JGK20}, 
pressure from infrared (IR) radiation created through the reprocessing 
of the starlight by dust grains 
\citep[e.g.][]{Thompson05,Murray10,Skinner15}, 
and the direct input of mechanical energy in the form of stellar winds
\citep[e.g.][]{Avedisova72,Castor75,Weaver77,KooMcKee92a,KooMcKee92b,Vink01}. 

The importance of these mechanisms has been debated in the 
recent literature, with several observational studies attempting to 
assess the relative contributions. This includes observations of evolved 
clusters in the Large Magellanic Cloud (LMC) by 
\citet{pellegrini11,Lopez11, Lopez14, McLeod19}, in the Milky Way 
\citep{Rosen14}, in nearby `normal' galaxies \citep{McLeod20,chevance20}, 
and in nearby galaxies with more extreme star-forming environments 
\citep{Levy20}. The recent works of \citet{Olivier20} and \citet{Barnes20} 
have sought to make the same evaluation but at very early times (in deeply 
embedded clusters) in the Milky Way. 

Prior to these recent empirical studies inter-comparing different feedback 
effects, earlier observations  suggested that the X-ray luminosities of 
nebulae were too low to be explained by standard wind models 
\citep{Dunne03,Townsley03,Townsley06,Townsley11}. Other works sought 
to explain this deficit 
\citep{GarciaSegura96,CapriottiKozminski01,Mackey15,ToalaArthur18,ElBadry19}, 
suggesting turbulent mixing and radiative cooling as sinks of energy, 
and \citet{HCM09} appealing to the leakage of hot wind gas out of the 
cloud interior due to turbulence-induced porosity.

In this paper, we present a model for the evolution of bubbles driven by stellar 
winds in the presence of strong interface cooling, induced by turbulent mixing.
Turbulent motions in the hot gas are derived from the kinetic energy of the wind, 
with Kelvin-Helmholz and other  instabilities growing at interfaces with the 
dense cloud material, itself highly structured due to background turbulence.  
Motivated by recent work on turbulent cooling-mixing layers in the context of 
multiphase galactic winds 
and the circumgalactic medium \citep{GronkeOh18,FieldingFractal20,Tan20}, we argue 
that this cooling can be very strong, removing up to 99\% of the injected energy. 
The cooling is strong enough to make the dominant phase of the bubble evolution 
momentum-driven (i.e. $p\propto t$). Our model therefore differs significantly 
from the well-known \citet{Weaver77} solution, which is energy-driven ($E\propto t$, 
with energy conserved interior to the bubble). Building on previous models for 
momentum-driven winds \citep{Steigman75}, we derive predictions for the evolution 
of important quantities such as the bubble's size and expansion rate, the momentum 
carried by the swept-up gas exterior to the bubble, and the energy and pressure 
interior to the bubble. 
Our goals are to characterize the dependence of bubble evolution on input 
wind power and ambient cloud properties, and to explain the physical mechanisms 
controlling the evolution.

In a companion paper \citep[][hereafter, Paper II]{PaperII} we present 
numerical simulations to validate the theory we develop here, and to 
constrain its few free parameters. We also discuss past 
numerical studies that have  explored this subject.

We have, by design, simplified the problem under consideration.  Our 
model ignores the effects of magnetic fields and extended star formation (in 
both space and time). Real astronomical systems of course have multiple sources 
of energy and additional physical elements (such as magnetic fields and strong 
stellar radiation). By tackling a simpler problem, however, we are able to 
develop a theoretical framework and make quantitative predictions that can 
be applied to interpret observations of star-forming clouds. We regard this as 
a valuable first step towards comprehensive understanding of the role of winds 
within the array of feedback processes.

The structure of this paper is as follows. In \autoref{sec:theory} we lay 
out the details of our theory for evolution of wind-driven bubbles in the 
turbulent ISM. In \autoref{sec:applications} we provide a guide to applying 
our theory in interpreting observations.  Finally, we place our work in 
context in \autoref{sec:discussion}, and conclude with a summary of our 
findings in \autoref{sec:conclusion}.

\section{Theory}
\label{sec:theory}

In this section we briefly review previous analytic theory of stellar 
wind bubble evolution and then provide an in-depth explanation of our 
new theory.

\begin{figure*}
    \centering
    \includegraphics[width=\columnwidth]{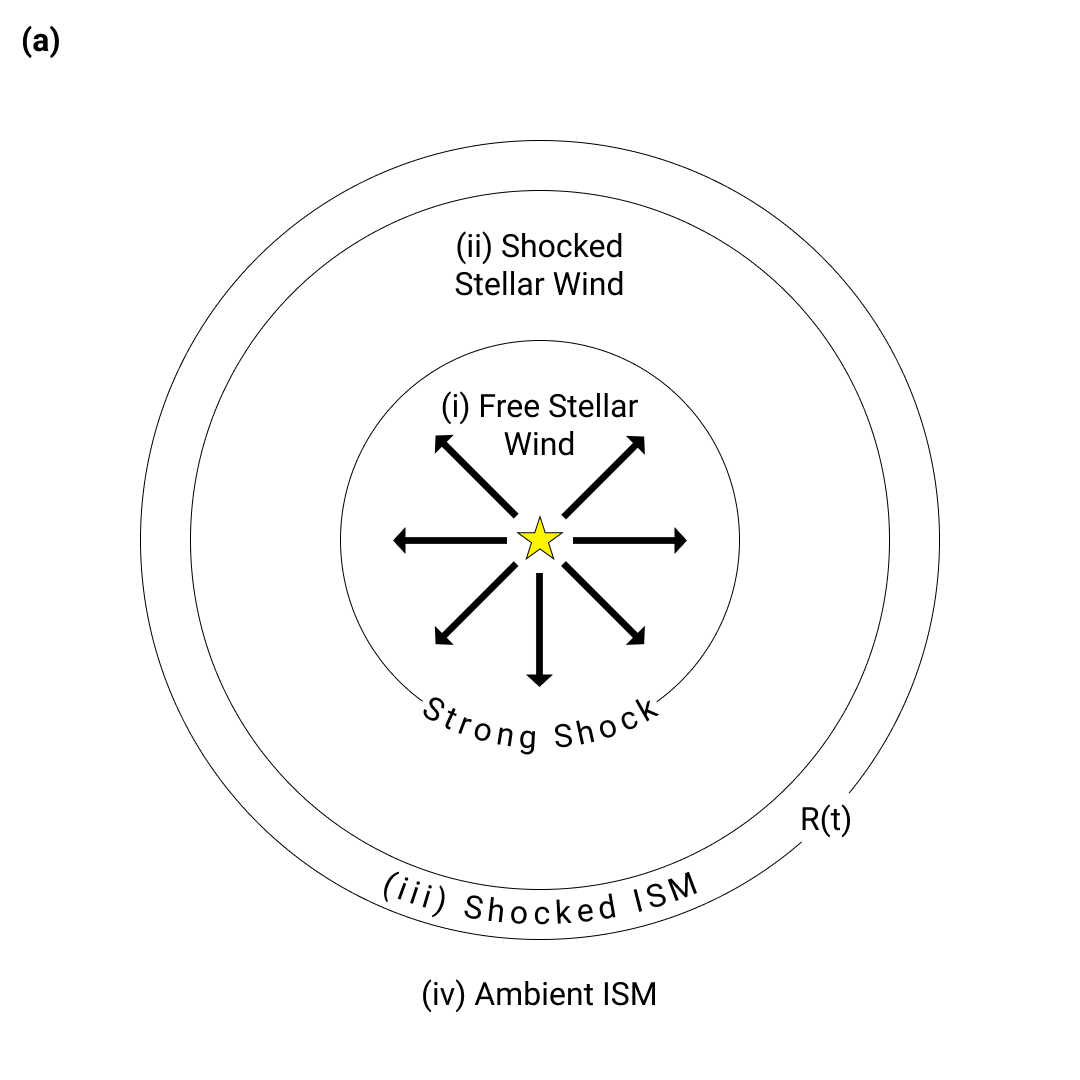}
    \includegraphics[width=\columnwidth]{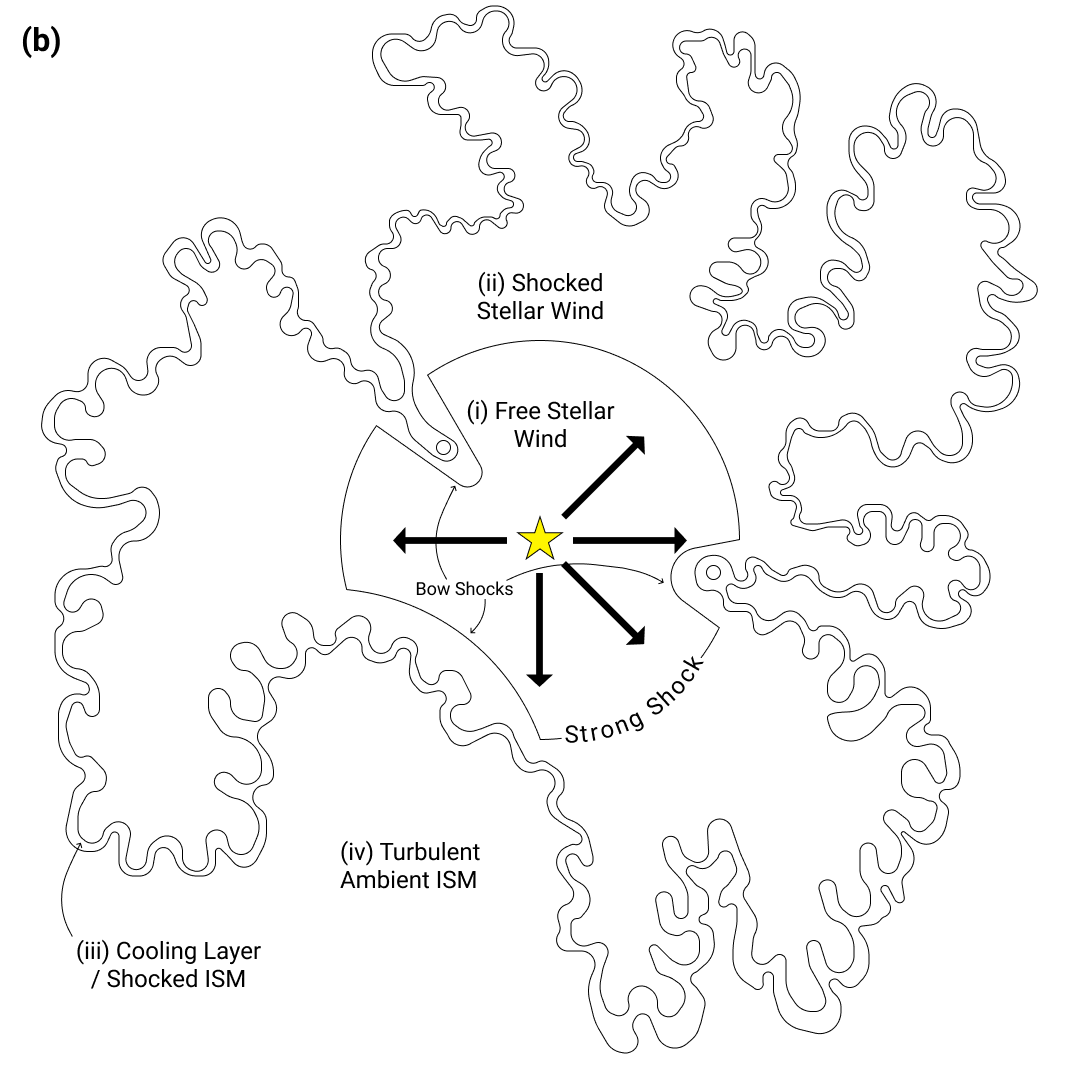}
    \caption{Schematic diagrams illustrating the structure of an 
    expanding wind-driven bubble for (a)  the classical pressure-driven 
    bubble of \citet{Weaver77}, 
    with a uniform background and cooling only at the leading shock; and (b) 
    the case explored in the present work, where the background is a turbulent
    medium and there is efficient  cooling at the interface between the wind 
    and the shell.  For both, the innermost region (i) is the same, consisting 
    of a free-flowing (hypersonic) radial stellar wind. In  both cases, there 
    is strong shock in which much of the wind's kinetic energy is thermalized; 
    outside of this is region (ii), consisting of hot, shocked stellar wind. 
    Both phases of the stellar wind have low density and minimal cooling.
    In (b), dense clumps of gas create bow shocks which increase the overall
    obliquity of the shock front, so that the post-shock radial velocity 
    remains higher.  In (a), there is a contact discontinuity (and an 
    evaporative flow, if conduction is included) at the interface between 
    the shocked wind and the surrounding dense shell.  In (b), the shocked 
    wind flows into the dense shell, interacting strongly in a turbulent 
    interface layer, which has a fractal structure.  The 
    intermediate-temperature mixture of hot shocked wind gas and dense shell 
    gas cools rapidly and merges into the shell. In both (a) and (b), most of 
    the mass in the shell consists of shocked ISM gas that has cooled down.
    The outermost region (iv) is ambient ISM gas, uniform in (a) but turbulent 
    and inhomogeneous in (b). Image credit: Cameron Lancaster.}
    \label{fig:schematics}
\end{figure*}

\subsection{Classical Stellar Wind Bubble}
\label{subsec:classic_wind}

The classical solution for the structure and evolution of the bubble produced 
by a constant luminosity wind expanding into a uniform background medium was 
described by \citet{Weaver77} 
\citep[see also e.g.][]{Avedisova72,Castor75,McCrayKafatos87}. We focus on 
the pressure driven (PD) stage of evolution. This stage was traditionally 
thought to characterize most of the evolution when a wind-blown bubble is
produced in a star-forming cloud by a single massive star or cluster. 
In this evolutionary stage, energy injected by the wind builds up within 
the expanding bubble's interior, stored in hot gas that is highly over-pressured 
relative to the ambient surrounding cloud.  As it expands into the ambient 
medium, the over-pressured bubble performs work on the background medium,
accelerating ambient gas and collecting it in a dense shell surrounding 
the low-density bubble interior.  Provided that the bubble expansion is
supersonic with  respect to ambient gas (always  true for the cold 
gas in giant molecular clouds [GMCs]), the ambient gas is initially shocked 
to high temperature, then rapidly cools down to condense into the shell that 
surrounds the hot bubble interior.  

\citet{Weaver77} divide the structure of the solution into four distinct regions: 
\begin{itemize}
    \item[(i)] Free Hyper-Sonic Wind
    \item[(ii)] Shocked Stellar Wind
    \item[(iii)] Shocked, Cooled Shell of Interstellar Gas
    \item[(iv)] Ambient Interstellar Gas;
\end{itemize}
this structure is laid out schematically in panel (a) of 
\autoref{fig:schematics}.

Under the assumption that all of the thermal energy created in the leading 
shock is radiated away while the bubble  interior remains non-radiative, 
the similarity solution for the bubble radius is
\begin{equation}
    \label{eq:rbub_weaver}  
    R=\RWeaver (t) \equiv 
    \left(\frac{125}{154 \pi}\right)^{1/5} 
    \left(\frac{\Lwind t^3}{\rhobar} \right)^{1/5},
\end{equation}
where $\Lwind$ is the wind luminosity, $\rhobar$ is the mass density 
of the background, and $t$ the time.  We use the ``W'' subscript to denote 
the Weaver solution. The dimensionless prefactor is approximately $0.76$. 
During a short-lived earlier stage before the shocked ambient gas cools 
(analogous to the Sedov-Taylor stage of a supernova remnant), the self-similar 
solution for the outer radius has the same form, but with a dimensionless 
prefactor $0.88$. 

The pressure in the shocked stellar wind evolves as 
\begin{equation}
    \label{eq:pbub_weaver}
    P=\PWeaver (t) \equiv \frac{5}{22\pi} 
    \left(\frac{125}{154\pi}\right)^{-3/5}
    \left(\frac{\Lwind^2 \rhobar^3}{t^4} \right)^{1/5} \, .
\end{equation}
In the PD bubble solution, the radial momentum carried by the 
shell\footnote{Throughout this work we will treat the total radial momentum
of a given solution as synonymous with the radial momentum in the swept up 
shell, as the momentum in the bubble interior is negligible.}
is given by
\begin{equation}
    \label{eq:prbub_weaver}
    \pr = \prWeaver (t) \equiv \frac{4\pi}{5} \left( \frac{125}{154 \pi}\right)^{4/5}
    \left( \Lwind^4  \rhobar t^7 \right)^{1/5} \, .
\end{equation}
Of the total wind energy $\Lwind t$ emitted up to time $t$, nearly 
half ($45\%$) is stored in thermal energy in the bubble interior:
\begin{equation}
    \label{eq:eth_weaver}
    E_{\rm th, W} = \frac{3}{2} \PWeaver \VWeaver
    = \frac{5}{11}\Lwind t \, , 
\end{equation}
where $\VWeaver$ is the volume of the bubble in this solution. Meanwhile 
a much smaller fraction ($19\%$) has gone into the kinetic energy of 
the swept-up shell
\begin{equation}
    \label{eq:ekin_weaver}
    E_{\rm kin,W} = \frac{1}{2} \Msh 
    \left(\frac{\diff \RWeaver}{\diff t} \right)^2
    = \frac{15}{77} \Lwind t
\end{equation}
where $\Msh = 4\pi \rhobar R^3 /3$ assuming all the swept-up gas is 
concentrated in the shell. The remaining $27/77 \approx 35\%$ of the 
wind energy is presumed to be radiated away from ISM gas that was 
shocked as it was swept into the advancing front of the bubble, and 
then cooled efficiently at high density. It is important to note that 
in the simplest version of the PD bubble solution, radiative cooling 
\textit{only} occurs for post-shock, swept-up ISM gas.  However, 
\citet{Weaver77} do also discuss late-stage evolution when the interior 
of the bubble drops to low enough temperature that it becomes radiative.  

\subsection{Pressure-Driven Bubble with Interface Cooling}
\label{subsec:elbadry_wind}

It is widely appreciated that it is difficult to maintain an idealized 
contact discontinuity, as exists in the classical PD bubble solution, between 
hot, diffuse gas in the bubble interior (shocked wind) and cool, dense gas of 
the shell surrounding it (swept-up ISM) 
\citep[e.g.][]{GarciaSegura96,CapriottiKozminski01}.  Turbulent motions at 
this interface, arising from ISM turbulence or from nonlinear development of
instabilities 
\citep[e.g.][]{Vishniac83,Vishniac94,VishniacRyu89,Blondin98,Bucciantini04,FoliniWalder06, Ntormousi11,Michaut12,Sano12,Pittard13,Badjin16}, 
will mix together the hot and cool gas. Given the efficiency 
of cooling at the resulting intermediate temperatures, most of energy 
carried into the interface by the hot gas would be radiatively cooled away. 

\citet{ElBadry19} used this fact to derive a simple modification to the 
PD bubble evolution for the case in which a fraction $\theta$ of the wind 
luminosity is lost to radiation due to turbulent mixing at the interface 
between the hot interior and the dense shell of the bubble:
\begin{equation}
    \label{eq:theta_def}
    \theta \equiv \frac{\lint}{\Lwind} \, ,
\end{equation}
where $\lint$ denotes the total radiative loss from the interface. 
Allowing for these losses, 
and assuming that $\theta$ is constant in time, the overall bubble evolution
follows the same form as in \autoref{eq:rbub_weaver} - \autoref{eq:ekin_weaver} 
with the substitution $\Lwind \to (1 -\theta)\Lwind$. This solution has the 
same dependence that the Weaver solution has on all of the relevant
parameters.  For example, the shell radius and outward radial momentum would follow 
\begin{subequations}
    \begin{align}
    \label{eq:Rb_EB}
    \REB(t) \equiv \left(\frac{125}{154\pi} \right)^{1/5}
    \left(\frac{\Lwind (1-\theta)t^3}{\rhobar} \right)^{1/5}\\
    \label{eq:pr_EB}
    \prEB (t) \equiv \frac{4\pi}{5} 
    \left( \frac{125}{154 \pi}\right)^{4/5}
    \left[(1-\theta)^4 \Lwind^4 \rhobar t^7 \right]^{1/5}
    \, .
    \end{align}
\end{subequations}
We note that physically, the bubble expansion is still assumed to be 
primarily driven by the thermal pressure of the interior, just at a 
reduced rate. 

The interface cooling, $\lint$, is quite distinct from the energy that 
is lost to radiation as shocked ISM gas cools down and joins the outside 
of the shell.  In the \citet{ElBadry19} solution, the radiative losses at 
the leading shock become $(27/77)(1-\theta) \Lwind$, i.e. the same fraction 
of ``available'' energy is still lost to cooling in the swept-up, shocked 
ISM (region (iii) in \autoref{fig:schematics} panel $a$). To make this 
distinction more explicit we will define the total fraction of energy that 
is lost to radiative cooling as 
\begin{equation}
    \label{eq:tot_theta_def}
    \Theta \equiv \frac{\ecool}{\Lwind} \, .
\end{equation}
For the \citet{ElBadry19} solution, 
$\Theta = 0.35 + 0.65 \theta$ 
and thus\footnote{Here, we use $0.65$ (0.35)  
as approximate shorthand for $50/77$ ($27/77$).} 
$1-\Theta = 0.65\left( 1-\theta \right)$.

\citet{ElBadry19} evaluated $\theta$ in the case where the mixing is 
governed by an effective turbulent diffusivity 
$\kappaeff = \lambda \delta v$, where the right-hand side 
represents the product of a spatial ($\lambda$) and velocity ($\delta v$) 
scale of turbulence in the  mixing layer. Since the simulations were 
spherical, $\kappaeff = \lambda \delta v$ was treated as an 
arbitrary parameter, and a range of values were explored.

Based on a combination of predicted scaling relations and numerical 
measurements, \citet{ElBadry19} showed that $\theta$ depends on  the 
ambient medium density $\rhobar$ and cooling rate $\Lambda(T_\mathrm{pk})$ 
at the temperature of maximal cooling, $T_\mathrm{pk}$, as 
\begin{equation}\label{eq:thetasoln}
    \frac{\theta}{1-\theta} \approx (\kappaeff \rhobar)^{1/2} \left[\frac{\Lambda(T_\mathrm{pk})}{(k_B T_\mathrm{pk})^2 }  \right]^{1/2} 
\end{equation}
where $k_B$ is the Boltzmann constant.

The pressure is relatively constant in a mixing/cooling interface, but 
the temperature varies.  It is useful to define the minimum cooling time as
\begin{equation}\label{eq:tcool}
    \tcool \equiv \frac{P}{n^2 \Lambda(T_\mathrm{pk} )} 
    =  \frac{(k_B T_\mathrm{pk})^2}{P\Lambda(T_\mathrm{pk})} \, ,
\end{equation}
It is straightforward to show that \autoref{eq:thetasoln} is equivalent to 
$\theta/(1-\theta)\approx 1.1 (\kappaeff/t_\mathrm{cool})^{1/2}/\dot{R}_{\rm EB}$ 
for the modified PD bubble solution.

We further note that the \citet{ElBadry19} simulations included classical 
microphysical (``Spitzer'') thermal conduction in addition to turbulent 
diffusivity.  As discussed in that paper \citep[following][]{Weaver77}, nonzero 
thermal conduction leads to evaporation of gas from the interface into the hot 
interior, which results in an increase of the hot gas mass and a decrease of its 
temperature (with pressure unchanged).\footnote{
Ablation (by conduction or Kelvin-Helmholz instabilities) of clouds that have been shocked and overtaken by the expanding high-velocity wind bubble can also add mass to the hot gas \citep[e.g.][]{CowieMcKee77,Klein94,Scannapieco2015,Schneider2017,Gronke2018}.}  
The thermal conduction does not lead to significant cooling because the conduction is only important at high temperatures when cooling is weak.  The cooling that does occur is a result of turbulent mixing.

\subsection{The Efficiently Cooled (EC) Solution}
\label{subsec:ecw_theory}

We now suppose that the energy losses in the bubble/shell interface are
so extreme that the PD regime is no longer relevant. A schematic of this 
scenario is given in \autoref{fig:schematics} panel $b$. In this limit, the 
mixing (and subsequent radiative cooling) of the shocked wind and dense shell 
gas is extremely efficient, so that little or no energy builds up in the interior 
of the bubble. When cooling is maximal, only momentum conservation need be 
considered. We adopt the term ``efficiently cooled'' (EC) to describe the solution 
in this limit. 

Given a constant mechanical luminosity $\Lwind$ and mass loss rate $\mdot$, 
the wind has asymptotic velocity $\vw$ and momentum input rate 
$\pdot$ given by
\begin{equation}
    \vw \equiv \left({\frac{2\Lwind}{\mdot}}\right)^{1/2} \, 
\end{equation}
and
\begin{equation}\label{eq:pdot}
    \pdot \equiv \vw\mdot = 2 \Lwind/\vw = \left({2\Lwind\mdot}\right)^{1/2}  \, .
\end{equation}
These parameters fully determine the fluid variables in the 
free wind (section (i) of \autoref{fig:schematics} panel b). For the shocked wind (section (ii) of \autoref{fig:schematics} panel b), in 
\autoref{app:wind_interior} we assume a steady subsonic radial flow to evaluate the fluid variables.

In the limit that no energy is able to build up within the wind bubble, 
the momentum evolution of the shell surrounding the bubble is entirely 
determined by the input momentum $\pdot$. In reality, the actual momentum 
can be somewhat larger than the input wind momentum, though we expect these 
deviations to be small. We parameterize any enhancement via a ``momentum enhancement 
factor'' $\alpha_p$ (the fundamental free parameter of our theory), and 
write the momentum equation as
\begin{equation}
    \label{eq:momentum_ecw}
    \frac{\diff}{\diff t} \left(\Msh 
    \langle v_r\rangle \right) = \alpha_p \pdot 
\end{equation}
where $\Msh$ is the mass swept into a shell by the expanding bubble
from the ambient ISM and $\langle v_r\rangle$ is the mass-weighted 
average radial velocity of that swept-up gas. Assuming that $\alpha_p$ 
and $\pdot$ are constant in time, the radial momentum of the shell is
therefore
\begin{equation}
    \label{eq:pEC}
    \pr = \alpha_p p_{\rm EC} \equiv \alpha_p \pdot t 
    =2\alpha_p \frac{\Lwind}{\vw}t \, ,
\end{equation}
where $p_{\rm EC}$ is the result for the case when $\alpha_p=1$, when 
cooling is maximally efficient.

It is reasonable to assume, given the statistical homogeneity of the 
turbulent background mass density field, that the swept-up mass is
approximately the product of the volume of the bubble and the mean 
background density
\begin{equation}
    \label{eq:mass_assumption}
    \Msh \approx \Vbub \rhobar \, .
\end{equation}

Given the statistical isotropy of the background turbulence, it 
is also reasonable to assume that $\langle v_r\rangle$ is mainly 
dependent on the rate of change of the bubble volume and not on 
its specific geometry. To this end we define the bubble's 
``effective radius" as
\begin{equation}
    \label{eq:reff_def}
    \reff \equiv \left(\frac{3\Vbub}{4\pi} \right)^{1/3}
\end{equation}
and assume that 
\begin{equation}
    \label{eq:velocity_assumption}
    \langle v_r\rangle \approx \frac{\diff \reff}{\diff t}   \, . 
\end{equation}
The deviation from equality in \autoref{eq:mass_assumption} and 
\autoref{eq:velocity_assumption} are essentially geometric and mostly 
driven by the inhomogeneity and anisotropy of the surrounding gas. For 
example, the bubble might preferentially expand in directions where the 
ambient gas has lower density, so that $M_{\rm sh} < \Vbub \bar\rho$.  

We now rewrite \autoref{eq:momentum_ecw} as 
\begin{equation}
    \label{eq:momentum_ecw2}
    \frac{\diff }{\diff t} \left( \frac{4\pi}{3} \rhobar 
    \reff^3 \frac{\diff \reff}{\diff t}\right) = \alpha_R \alpha_p
    \pdot 
\end{equation}
where $\alpha_R$ is an order-unity parameter that accounts 
for any departure from equality in \autoref{eq:mass_assumption} and 
\autoref{eq:velocity_assumption}. 
We note that, unlike the case for other parameters related to energy 
(see below), there is not a direct relationship between $\alpha_R$ and 
$\alpha_p$. However, as we show in Paper II, $\alpha_R$ is very close to 1 
for our simulated bubbles. Taking $\alpha_R$ to be approximately constant, 
\autoref{eq:momentum_ecw2} can be integrated to obtain
\begin{equation}
    \label{eq:rbub_ecw}
    \reff (t)  = 
    \left(\frac{3\alpha_R \alpha_p}{2\pi} 
    \frac{\pdot t^2}{\rhobar} \right)^{1/4} 
    \equiv (\alpha_R\alpha_p)^{1/4} \REC \, .
\end{equation}
Here, $\REC$ denotes the solution for an exactly spherical bubble 
expanding in an uniform ambient medium when the shell momentum is 
taken to increase at a constant rate $\pdot$, as has been derived by many authors 
\citep[e.g.][]{Steigman75,KooMcKee92a,KooMcKee92b,KimOstrikerRaileanu17}; 
this would apply in the limit of no energy build-up within the bubble 
interior.

We now wish to determine the energetics of the bubble interior. In 
\autoref{app:wind_interior} we give a derivation of the expected 
values for the fluid variables in the bubble interior under the 
assumption of spherical geometry with a  perpendicular shock 
dividing the free wind from the post-shock sub-sonic, steady, radial 
downstream flow.

Under these assumptions the energy of the bubble interior can be expressed as 
\begin{equation}
    \label{eq:ebub_relation}
    \Ebub = \frac{1}{2}\pdot \ratr \reff  = (\alpha_R \alpha_p)^{1/4} 
    \ratr E_{b,{\rm EC}} \, .
\end{equation}
Here, $E_{b,{\rm EC}}= \pdot R_{\rm EC}/2$ is the bubble energy in the case 
where the bubble is occupied solely by the free wind, with $\ratr=\alpha_p = 1$. 
More generally, $\ratr$ encodes the extra energy contained within the bubble 
($\ratr > 1$ when $\alpha_p >1)$, with  \autoref{eq:S_derive} and 
\autoref{eq:RbRfsoln} relating $\ratr$ to $\alpha_p$.  $\ratr$ is approximately 
linear in $\alpha_p$, with $\ratr \approx \alpha_p$ within 6\%.

The pressure in the shocked wind is nearly constant, equal to its immediate 
post-shock value, $3 \pdot /(16 \pi \rfree^2)$, where $\rfree$ is the outer 
radius of the free wind region, defined analogously to \autoref{eq:reff_def}.  
Written in terms of $\reff$ and $\alpha_p$, this is
\begin{equation}
    \label{eq:pressure_time}
    \Pbub = \frac{3\pdot}{16\pi \reff^2}
   \left[ \frac{2}{3}\alpha_p + \left( \left(\frac{2}{3}\alpha_p\right)^2 - \frac{1}{3}\right)^{1/2}\right],
    %\left(\frac{\reff}{\rfree} \right)^2\,.
\end{equation}
where the term in square brackets is $\approx (3 \alpha_p -1)/2$ within $4\%$ 
for the range $1\le \alpha_p \le 4$ we find in Paper II.

We now consider the kinetic energy of the shell driven by the bubble. If 
we assume that the vast majority of the input momentum at any time is stored 
in the shell, we may write its radial kinetic energy as:
\begin{equation}\label{eq:EKr}
    \Ersh = \frac{\pr^2}{2\Msh} 
    = \frac{3\left(\alpha_p \pdot t\right)^2}{8\pi \rhobar \reff^3}
    = \frac{\alpha_p }{4\alpha_R}\pdot\reff .
\end{equation}
The ratio of the bubble interior's energy to the radial kinetic energy of 
the shell is only weakly dependent on time as
\begin{equation}
    \label{eq:kin_bub_ratio}
    \frac{\Ebub}{\Ersh} = 2\frac{\alpha_R}{\alpha_p}
    \ratr \, ,
\end{equation}
which should be approximately 2 
since $\ratr \approx \alpha_p$ and $\alpha_R\sim 1$. This can be 
compared to the ratio between \autoref{eq:eth_weaver} and 
\autoref{eq:ekin_weaver}, which is equal to $7/3=2.3$.  Thus, even though 
the EC solution has two (or more) orders of magnitude lower energy than the 
Weaver solution because most energy is radiated away, the ratio of interior 
to shell energy is comparable.  

In addition to expanding radially, the shell may also acquire turbulent 
motion, such that $\Esh=\Ersh+ \Etsh$ is its total kinetic energy. 
This shell turbulence is a side-effect of the interface instabilities that 
induce mixing and cooling.  There is no \textit{a priori} prediction 
for the turbulent energy, but we can describe the level of turbulent energy 
relative to the radial kinetic energy as
\begin{equation}
    \label{eq:fturb_def}
    f_{\rm turb} \equiv \frac{\Etsh}{\Ersh} \, .
\end{equation}
With this definition, the total energy of the shell plus bubble interior
(neglecting the small thermal energy of the shell) will be 
$\Esh + E_b = ( 1 +  f_{\rm turb} + 2\alpha_R \ratr/\alpha_p )\Ersh$. 

We can use the above to formulate an expression for the fraction of the energy input rate  
lost to cooling,  $\Theta$.  From conservation of energy, the energy that is 
\textit{not} lost to cooling is $\Lwind \int (1-\Theta) dt$, and this must be 
equal to $\Esh + E_b $. Using \autoref{eq:pdot} and \autoref{eq:EKr} and 
taking a derivative in time, we obtain 
\begin{equation}
    \label{eq:Theta_cool_prediction}
    1 - \Theta = 
    \left(\frac{1}{2} (1 + f_{\rm turb})\frac{\alpha_p}{\alpha_R} +\ratr \right)
    \frac{\dotreff}{\vw} \, .
\end{equation}
Evidently, because $\dotreff \propto t^{-1/2}$, the fraction of energy 
available after cooling is expected to be initially large, but decreasing in 
time.  Correspondingly, the fraction $\Theta$ of energy lost to cooling is 
expected to be initially small, but increasing with time.  
In the EC solution, the bubble energy increases as 
$E\propto \reff \propto t^{1/2}$; this may be contrasted with $E\propto t$ 
in classical wind-driven bubble solutions.

Since the dimensionless prefactor in \autoref{eq:Theta_cool_prediction} is 
order-unity, the fraction of energy retained ($1-\Theta$) is comparable to the 
ratio of the shell expansion velocity to the original wind velocity 
($\dotreff/\vw$), which we show (in Paper II) drops to $\sim 1\%$ or lower over 
time for bubbles expanding in the dense environments of star-forming clouds.

\begin{figure*}
    \centering
    \includegraphics{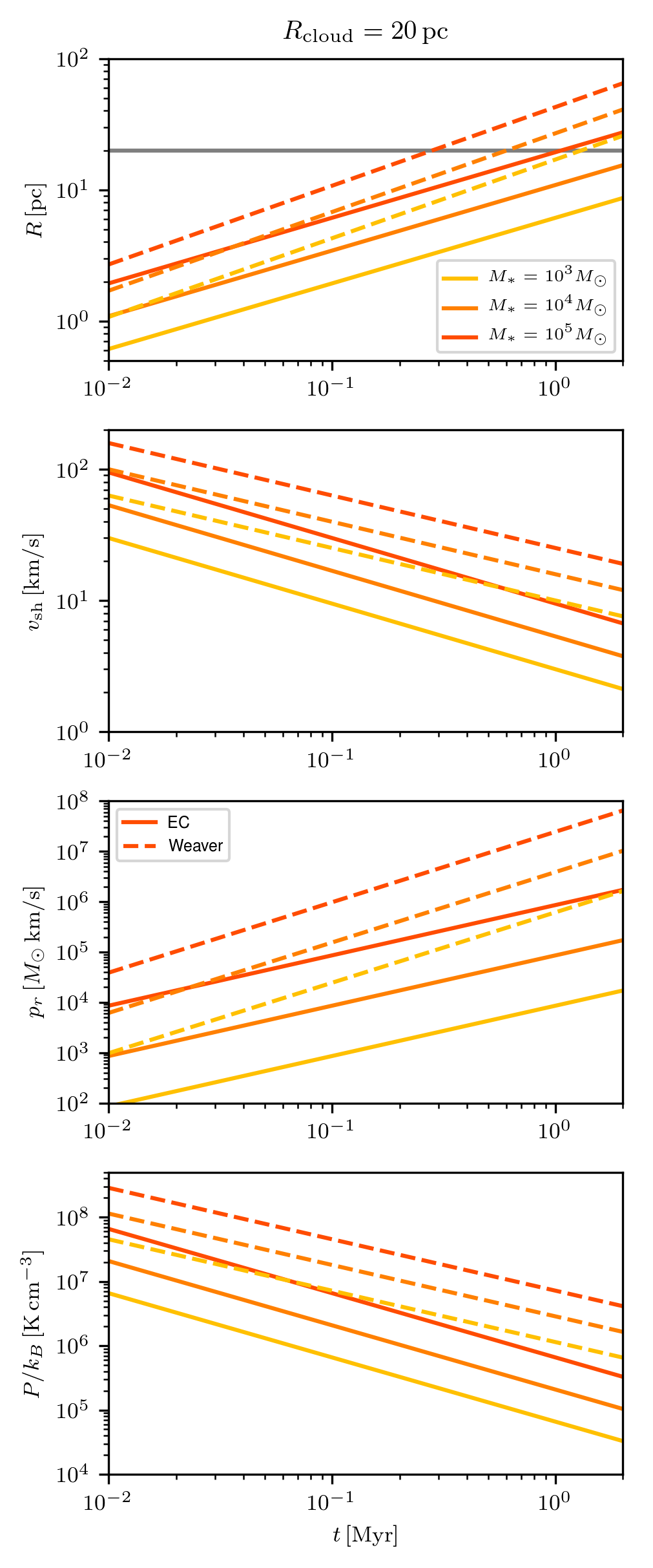}
    \includegraphics{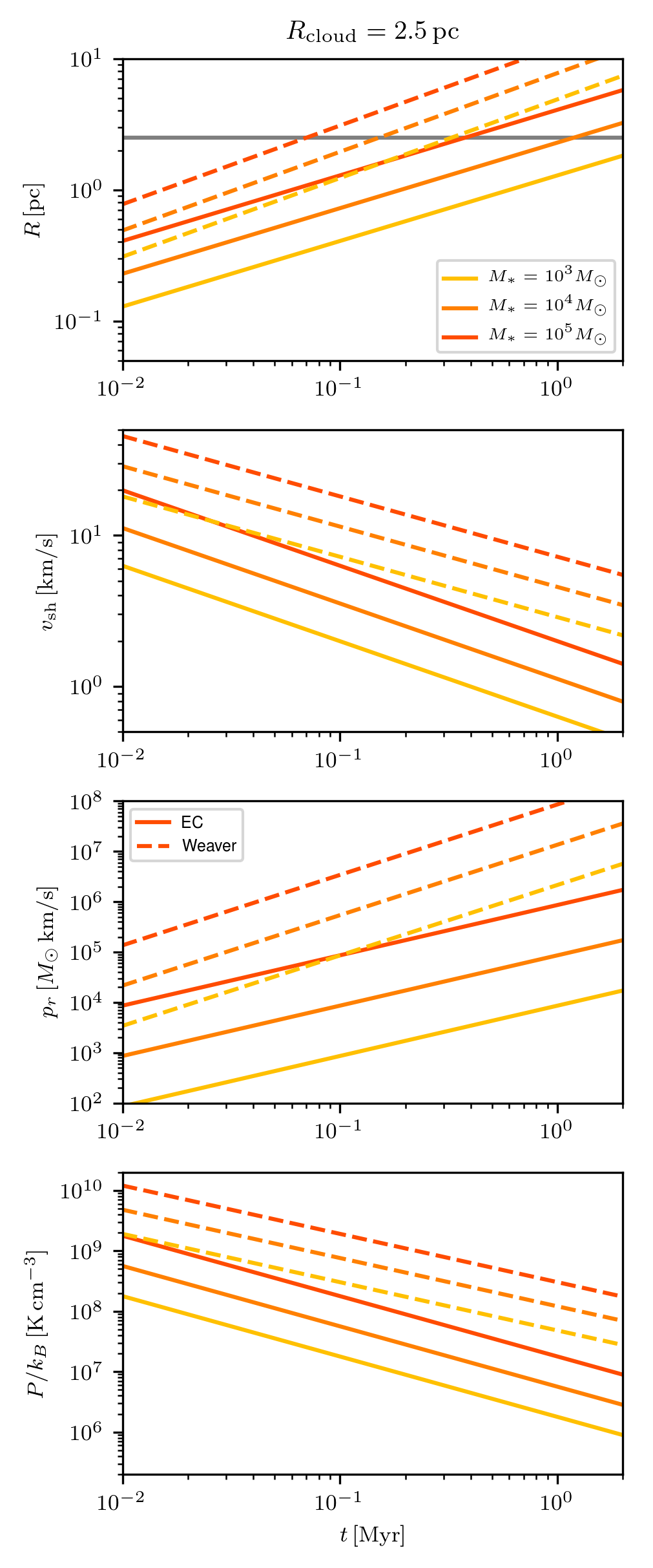}
    \caption{
    Temporal evolution of physical quantities for a wind-driven bubble in both 
    the Weaver theory (dashed lines) and the theory presented here (solid lines). 
    From top to bottom we show the bubble radius, the velocity and radial momentum 
    of the swept-up gas surrounding the bubble, and the thermal pressure in shocked 
    wind in the bubble interior.  We consider a cloud with mass 
    $\mcloud = 10^5 M_{\odot}$ and a radius  of $R_{\rm cloud}= 20\, {\rm pc}$ 
    (left panels) or $R_{\rm cloud}= 2.5\, {\rm pc}$ (right panels). The horizontal 
    grey line in each of the top panels indicates the cloud radius. For every value 
    we explore three different values of the mass of the star cluster driving the 
    wind, $M_* = 10^3, 10^4, 10^5\, M_{\odot}$ in yellow, orange, and red respectively. 
    }
    \label{fig:tevol_weaver_comp}
\end{figure*}

Having explained the evolution of the main physical quantities in the EC theory, 
we illustrate the significant differences between our theory and that of 
\citet{Weaver77} (reviewed in \autoref{subsec:classic_wind}) in 
\autoref{fig:tevol_weaver_comp}. Specifically, we show the evolution of the 
bubble's radius (or $\reff$ in our theory), the velocity, and radial momentum 
carried by the swept-up gas, and the pressure in shocked wind in the bubble's 
interior.  We show results for three different wind strengths (set by 
$M_*=10^3,\ 10^4,\ 10^5 M_\odot$), considering a cloud with mass 
$\mcloud = 10^5 M_{\odot}$ and radius $\rcloud = 20 \, {\rm pc}$ or 
$\rcloud = 2.5 \, {\rm pc}$.  The mean ambient density for the larger (smaller) 
cloud is $n_H=86\, {\rm cm}^{-3}$ ($n_H=4.4\times 10^4\, {\rm cm}^{-3}$). For 
all cases the central source has specific luminosity 
$\Lwind/M_* = 10^{34} \, {\rm erg\ s^{-1}\ M_{\odot}^{-1}}$ and specific mass 
loss rate of $\mdot/M_* = 10^{-2.5} \, {\rm Myr}^{-1}$, based on Starburst99 
\citep{SB99}. As shown below in \autoref{fig:wind_evol}, these values are 
appropriate for the first $\sim2.5\, {\rm Myrs}$ of star cluster evolution at 
solar metallicity. We have assumed $\alpha_R = \alpha_p= \ratr = 1$ for this 
comparison, so $\reff$ is equivalent to $\REC$.  The difference between our 
theory and that of \citet{Weaver77} is striking. In particular, for the radial 
momentum and interior pressure, the predictions differ by factors of 10-100.

\subsection{The Turbulent Diffusivity}
\label{subsubsec:EB_theta_eval}

We now consider expectations for the properties of the mixing layer and 
magnitude of the effective diffusivity $\kappaeff$. In general, a 
mixing/cooling interface layer of thickness $\Lmix$ must have a balance 
between diffusion of energy at volumetric rate $\sim \kappaeff P/\Lmix^2$ 
and radiative losses at volumetric rate $\sim n^2 \Lambda= P/\tcool$ in 
the mixed gas. For resolved interface layers with a range of $\kappaeff$, 
\citet{ElBadry19} numerically evaluated profiles and found 
$\Lmix \approx 2 (\kappaeff  \tcool)^{1/2}$, consistent with this 
expectation; this $\tcool$ is is based on the maximal cooling rate at a 
given pressure, defined in \autoref{eq:tcool}.  

If we consider a flow entering a mixing/cooling interface at velocity 
$\vrel$, it carries an enthalpy flux $(5/2)\vrel P$.  This flux 
must be balanced by the cooling per unit area, 
$n^2 \Lambda \Lmix =  P \Lmix/\tcool$.  Equating these rates, we obtain
$\vrel = (2/5)  \Lmix/\tcool \approx 0.8 (\kappaeff/t_\mathrm{cool})^{1/2}$.

What sets $\kappaeff$? A turbulent boundary layer would in general have 
a range of physical scales $\ell$, with corresponding velocities 
$\vt(\ell).$\footnote{Here, $\vt(\ell)$ refers to turbulence in the hot gas; 
\citet{FieldingFractal20} discusses the relation between turbulent amplitudes 
in the hot and cool (mixed) gas.} The spectrum of velocity fluctuations is 
usually assumed to follow a power law,
\begin{equation}
    \label{eq:vturb_strucf}
    \vt (\ell) = \vt(L) \left(\frac{\ell}{L} \right)^p
\end{equation}
where $p$ is the power-law index of this scaling, $L$ is the largest scale 
on which this turbulent structure function applies, and $\vt(L)$ is the
velocity at that scale.  

At scale $\ell$, the 
eddy-turnover (or flow-crossing) time is 
\begin{equation}
    \label{eq:te_deff}
    \te(\ell)\equiv \frac{\ell}{\vt(\ell)} 
    = \left(\frac{\ell}{L}\right)^{1-p}\frac{L}{\vt(L)}
    =\left(\frac{\ell}{L}\right)^{1-p}\te(L).
\end{equation}
Since $p<1$ in general, $\te(\ell)$ increases with $\ell$. 
Thus, if the cooling time is short enough that $\tcool <\te(L)=L/\vt(L)$, 
there always exists some scale $\lcool$ such that  
$\tcool =  \te(\lcool) = \lcool/\vt(\lcool)$ is satisfied. 
In the language 
of \citet{Tan20}, this condition is the same as saying that the 
Damk\"ohler Number is greater than 1, which is generally true 
in our simulations (see Paper II). 
As noted by \citet{FieldingFractal20}, it is 
this critical scale, which has $\lcool = \vt(\lcool) t_\mathrm{cool}$, that 
is most relevant for mixing-mediated cooling. Turbulent eddies at 
$\ell > \lcool$ will be too slow to directly mix gas in a way that enables 
rapid cooling, while eddies at $\ell < \lcool$ will simply further enhance 
mixing. Thus, the most relevant value for the effective turbulent 
diffusivity is
\begin{equation}
    \kappaeff  =  \lcool \vt(\lcool) = t_\mathrm{cool} [\vt(\lcool)]^2\, .
\end{equation}
Using this diffusivity, the thickness of the mixing/cooling layer would be 
$\Lmix\approx 2 \lcool$ and the inflow velocity of hot gas to the interface 
would be
\begin{equation}\label{eq:vrel}
    \vrel =0.8 \vt(\lcool)\approx \frac{\lcool}{t_\mathrm{cool}} \, . 
\end{equation}

The scale at which mixing/cooling occurs, and the characteristic velocity at 
that scale, can be written in terms of the turbulent properties at the 
energy-containing scale as 
\begin{equation}\label{eq:lcool}
\lcool = L \left[\frac{\vt(L)\tcool}{L}\right]^\frac{1}{1-p}    
\end{equation}
and
\begin{equation}\label{eq:vlcool}
    \vt(\lcool)= \vt(L) \left[\frac{\vt(L)\tcool}{L}\right]^\frac{p}{1-p}.
\end{equation}

The above describe expectations for the turbulent diffusion and the velocity 
of flow into the interface.  In addition to the inflow velocity, the rate 
of energy loss also depends on the area of the interface.  If the main 
energy-containing scale for the turbulence is $L$, the interface will
be irregular at scales $\lesssim L$. While turbulent mixing and cooling 
would still lead to inflow to the interface at $\vrel \sim \vt(\lcool)$, 
the net cooling is enhanced by having a surface that is highly corrugated 
at small scales. We turn to this in the next section.

\subsection{Fractal Nature of the Turbulent Boundary Layer}
\label{subsec:fractals}

In a recent investigation of turbulent, cooling boundary layers driven by 
Kelvin-Helmholtz instabilities between hot diffuse gas and cool dense gas, 
\citet{FieldingFractal20} pointed out that the surface of the interface obeys 
a fractal scaling law (see also \citealt{Tan20}, who connect this to results 
in the combustion literature).  \citet{FieldingFractal20} used this fractal 
nature to derive a prediction for the rate of mixing and cooling that occurs 
at shearing interfaces between cool, dense gas and hot, diffuse gas.

In particular, \citet{FieldingFractal20} argue that the rate of thermal 
energy loss to cooling will be equal to the rate at which energy can be 
simultaneously mixed and cooled into the turbulent interface, 
\[
    \dot E \approx \frac{5}{2} P \vt(\lcool) A({\lcool}) \, , 
\]
where $A({\lcool})$ is the area of the (fractal) interface at 
scale\footnote{\citet{FieldingFractal20} adopt the notation $w$ for the 
critical scale that here we denote $\lcool$.} $\lcool$.  The above corresponds 
to the same characteristic inflow velocity of hot gas 
$\vrel \approx \vt(\lcool) = \lcool/t_\mathrm{cool}$ as 
given in \autoref{eq:vrel},
while the total cooling rate takes into 
account the fractal area $A({\lcool})$ of the interface at the scale $\lcool$. 
The above relation is valid up to an order unity constant, which is quantified 
in \citet{FieldingFractal20} for the case where the interface is planar on large 
scales. The coefficient may depend somewhat on the details of the problem 
(including how turbulence is driven and large-scale geometry), so for present 
purposes we concentrate on scalings.

For the wind-blown bubble problem, we define an ``equivalent'' thermal energy 
flux  $\Phi_\mathrm{cool}$ that can be lost to cooling via interface mixing as 
the rate of energy loss divided by the surface area of an equivalent sphere,
\begin{equation}
    \label{eq:equiv_flux}
    \Phi_\mathrm{cool} \approx \frac{5}{2} P \vt(\lcool) 
    \frac{\Abub(r;\lcool)}{4 \pi r^2} \, .
\end{equation}
Here $\Abub(r;\lcool)$ is now the full fractal surface area of the 
wind-blown bubble when the bubble has some linear scale (for example, its
radius) $r$.

We will make the assumption that the thin interface between the shocked 
wind and the cool shell (i.e. between regions (ii) and (iii) in panel b 
of \autoref{fig:schematics}) can be described by a fractal of dimension 
$D>2$ (as it is an interface). We will refer to its ``excess dimensionality'' 
as $d \equiv D-2$.

Physically, this means that if the bubble has an overall linear scale $r$, 
the area of the bubble surface measured on scale $\ell$ will be
\begin{equation}
    \Abub(r;\ell) \approx 4 \pi r^2 \left(\frac{r}{\ell} \right)^d \, . 
\end{equation}

As in \autoref{subsubsec:EB_theta_eval}, we assume that the energy-containing 
scale of turbulence in hot gas near the interface is $L$ and that the 
turbulence can be described by a structure function of the form of 
\autoref{eq:vturb_strucf} for $\ell \le L$.  The power law index is 
expected to take on a value $p\sim 1/3$ because the turbulence in the hot 
gas is generally subsonic. The amplitude $\vt(L)$ likely depends primarily 
on instabilities at the interface since the wind is the main source of free 
energy, but the background cloud turbulence may be important in seeding them. 

Using \autoref{eq:lcool} and \autoref{eq:vlcool}, the equivalent energy flux 
is expected to follow
\begin{subequations}
\begin{eqnarray}
\label{eq:coolflux_a}
\Phi_\mathrm{cool} 
&\approx&
\frac{5}{2} P \vt(\lcool) \left(\frac{r}{\lcool} \right)^d \\
\label{eq:coolflux_b} 
&\approx&\frac{5}{2} P \vt(L) \left( \frac{\lcool}{L} \right)^{p-d} \left(\frac{r}{L} \right)^d  \\ 
&\approx& 
\frac{5}{2} P \vt(L) 
\left[\frac{\vt(L)t_\mathrm{cool} }{L} \right]^{\frac{p-d}{1-p}}
\left(\frac{r}{L} \right)^d 
\label{eq:coolflux_c}
\end{eqnarray}   
\end{subequations}
We emphasize that the fractal nature of the interface plays a major role 
in enhancing energy losses, given by  the factor $(r/\lcool)^d$ in 
\autoref{eq:coolflux_a}.  
It is important to note that strong inhomogeneity in a cloud (due to 
preexisting high Mach number turbulence) can make the fractal structure 
global, with $r\gg L$; from \autoref{eq:coolflux_c} this strongly 
enhances cooling. 

\citet{FieldingFractal20} measured $d\approx 1/2$ for the excess 
dimensionality of the fractal interface in the mixing/cooling layer of 
their simulations.  Combining this with $p=1/3$ for subsonic turbulence 
yields 
\begin{equation}
    \Phi_\mathrm{cool} \approx \frac{5}{2} P \vt(L) 
    \left[\frac{\vt(L)\tcool }{L} \right]^{-1/4} 
    \left(\frac{r}{L} \right)^{1/2}. 
\end{equation}
Since the \citet{FieldingFractal20} simulations were for a local rectangular 
box with a shear layer, rather than a global expanding bubble, in the situation 
they studied $r/L \rightarrow 1$.  The predicted scaling 
$\Phi_\mathrm{cool} \propto \vt(L)^{3/4} \tcool^{-1/4} $ with the 
measured large-scale turbulent velocity $\vt(L)$ and  imposed cooling time 
$\tcool$ were found to be in excellent agreement with numerical results.

For present purposes, in order to keep the relation to the spherical case more 
explicit and use the variables that we have already developed to represent the 
linear scale of the bubble, we will write the fractal area relation
\begin{equation}
    \label{eq:frac_area}
    \Abub(\reff;\ell) \equiv 4\pi \alpha_A \reff^2 
    \left(\frac{\reff}{\ell} \right)^d
\end{equation}
where $\alpha_A$ is an order unity constant, which we quantify in Paper II. 

Finally, we note that even if a properly-calibrated $\kappaeff= \vt(\lcool)\lcool$ 
were used and the cooling length were resolved in one-dimensional simulations, it 
would not be possible to capture the fractal nature of the interface and hence the 
total cooling would not be correctly captured.  To achieve the proper energy 
loss rate, an ``area-corrected'' effective diffusivity 
$\kappa_{\rm corr,1D}= \vt(\lcool)\lcool (r/\lcool)^{2d}$ would have to be 
adopted instead, where $r$ here now denotes the radius of the interface between the 
hot bubble and ambient medium in the 1D simulation.  From \autoref{eq:lcool}, this 
can be written in terms of the energy-containing scale of the turbulence as 
\begin{equation}\label{eq:kappa_corr}
    \kappa_{\rm corr,1D}=\vt(L) L 
    \left[\frac{\vt(L)\tcool}{L}\right]^{\frac{1+p-2d}{1-p}}
    \left(\frac{r}{L}\right)^{2d}\, .
\end{equation}

\subsection{Conditions for Efficient Cooling}
\label{subsec:cooling}

Here we analyze the conditions under which a wind 
can become ``efficiently cooled.'' One such condition could be stated 
as the point at which the momentum incurred by pressure work, given 
for the PD bubble with cooling by \autoref{eq:pr_EB}, is less than 
the direct momentum input from the wind, $\dot p_w t = p_{\rm EC}$
(\autoref{eq:pEC}, which is the total momentum in the EC case). 
This condition can be expressed formally as
\begin{equation}
    \label{eq:cooling_condition}
    1 - \Theta < 4\left(\frac{5}{6}\right)^{1/4}    \frac{\dot{R}_{\rm EC}}{\vw} \, .%
\end{equation}
Since ${R}_{\rm EC} \propto t^{1/2}$, the term on the right above 
varies as $t^{-1/2}$. If cooling losses are sufficiently weak, the 
efficient cooling condition may be satisfied only at the earliest 
times. If, however, cooling losses are strong, $ 1 - \Theta \ll 1$, 
essentially the whole bubble evolution may be in the efficient-cooling 
regime.  

The above gives us a necessary condition on $\Theta$ for the wind to 
be ``efficiently cooled.'' If an estimate of the cooling losses were 
known, then this test would decide whether the cooling-modified PD 
(\autoref{subsec:elbadry_wind}) or EC (\autoref{subsec:ecw_theory})
bubble solution is valid.  Since, however, we have no \textit{a priori} 
knowledge of $\Theta$ (except through \autoref{eq:thetasoln} and 
\autoref{eq:kappa_corr}), it is more useful to obtain a physically 
motivated condition for the EC solution to apply.

To this end, we compare the estimated thermal energy flux that can be lost 
to cooling at a turbulent interface, $\Phi_\mathrm{cool}$ 
(\autoref{eq:coolflux_c}), to the thermal energy flux that an unimpeded 
shocked wind would carry.  Immediately after the shock, at radius $\rfree$, 
this flux is $\Phi_\mathrm{w} \approx (5/2) P_{\rm ps} \vsw $ with 
$P_{\rm ps} = 3 \pdot/(16 \pi \rfree^2)$ $\vsw=\vw/4$ for a spherical shock. 
Since the pressure is approximately constant in the subsonic post-shock region 
while energy is conserved (see \autoref{app:wind_interior}), the energy flux 
carried into the boundary layer is 
$\Phi_\mathrm{w} \approx (5/2) P_{\rm ps} \vsw (\rfree/\reff)^2$.

The condition for efficient cooling is that the capacity for mixing/cooling 
matches or exceeds the rate at which thermal energy is advected to the 
interface ($\Phi_\mathrm{w}\lesssim\Phi_\mathrm{cool}$), which may be expressed as
\begin{equation}
\label{eq:veff_mixcool}
    \frac{2}{3\alpha_p-1}\vsw \lesssim \vt(L) 
    \left[\frac{\vt(L)t_\mathrm{cool} }{L} \right]^{\frac{p-d}{1-p}}
    \left(\frac{\reff}{L} \right)^d \, . 
\end{equation}
For the numerical coefficient on the left-hand side we have used the 
small-$\alpha_p$ approximation given in \autoref{eq:RbRf_approx}; when 
$\alpha_p \gg 1$ the prefactor would instead become $3/(4\alpha_p)$, from 
\autoref{eq:RbRfsoln}.  The exponent $(p-d)/(1-p)$ depends on the exact 
fractal dimension as well as the power of the turbulent structure function.  
If $d\approx 1/2$ and $p\approx 1/3$, as previously suggested by 
\citet{FieldingFractal20}, this exponent is $-1/4$, but in any case it is 
likely to be small and negative. Still, $t_\mathrm{cool}$ can be as  small 
as $\sim 10-100 \yr$ in dense clouds (see Paper II for numerical 
results). With typical 
$\vt(L) \sim 0.1 \vw \gtrsim 100 \kms$ and $L\sim 0.1 \reff\lesssim 1\, \pc$, 
the efficient-cooling condition is satisfied for a strong shock with 
$\vsw = \vw/4$.   Even if oblique shocks yield a larger hot-gas velocity 
$\vsw \lesssim \vw$, the efficient-cooling condition is likely to be at least 
marginally satisfied.

If we think of \autoref{eq:veff_mixcool} as representing an outer boundary 
condition on the flow within the bubble, it shows that lower turbulence levels 
will allow for larger $\alpha_p$, i.e. more buildup of shocked gas within the 
bubble.  Very high turbulence levels and large fractal dimension, in contrast, 
would lead to extremely efficient mixing, with $\alpha_p \sim 1$.

Finally, we remark that in numerical simulations, limited numerical resolution 
may impose a minimum resolved scale $\ellmin \sim \Delta x$, where  
$\lcool < \ellmin < L$. Following the arguments leading to 
\autoref{eq:coolflux_a}-\autoref{eq:coolflux_b} and then 
\autoref{eq:veff_mixcool}, mixing at $\ellmin$ would provide sufficient
cooling for the EC solution to be satisfied provided 
\begin{subequations}
\begin{eqnarray}
\label{eq:cool_min}
    \frac{2}{3\alpha_p-1} \vsw  &\lesssim& \vt(\ellmin) 
    \left(\frac{\reff}{\ellmin} \right)^d \\ 
    &\sim& \vt(L) \left(\frac{\ellmin}{L} \right)^{p-d} \left(\frac{\reff}{L} \right)^d\,.
\label{eq:cool_min_b}
\end{eqnarray} 
\end{subequations}
With $p-d \sim -1/6$, the cooling rate would tend to \textit{increase} 
as resolution and $\ellmin$ get smaller. Thus, if \autoref{eq:cool_min} 
is already satisfied for resolved turbulence at scale $\ellmin$ of a few 
$\Delta  x$, further improvement in the resolution would not alter the 
behavior given by \autoref{eq:rbub_ecw}. 

In Paper II we use \autoref{eq:cool_min} as an additional check whether 
we are in the efficiently cooled regime. However, due to resolution effects, 
the structure function of the turbulence is often steeper than it should be, 
meaning that cooling does not necessarily increase with decreasing scale 
\textit{all the way} to the resolution limit. If the true mixing rate is 
higher than the rate achieved in our simulations due to resolution limitations, 
the EC conditions in reality would be even  better satisfied, with $\alpha_p$ 
even closer to unity.
\begin{figure*}
    \centering
    \includegraphics{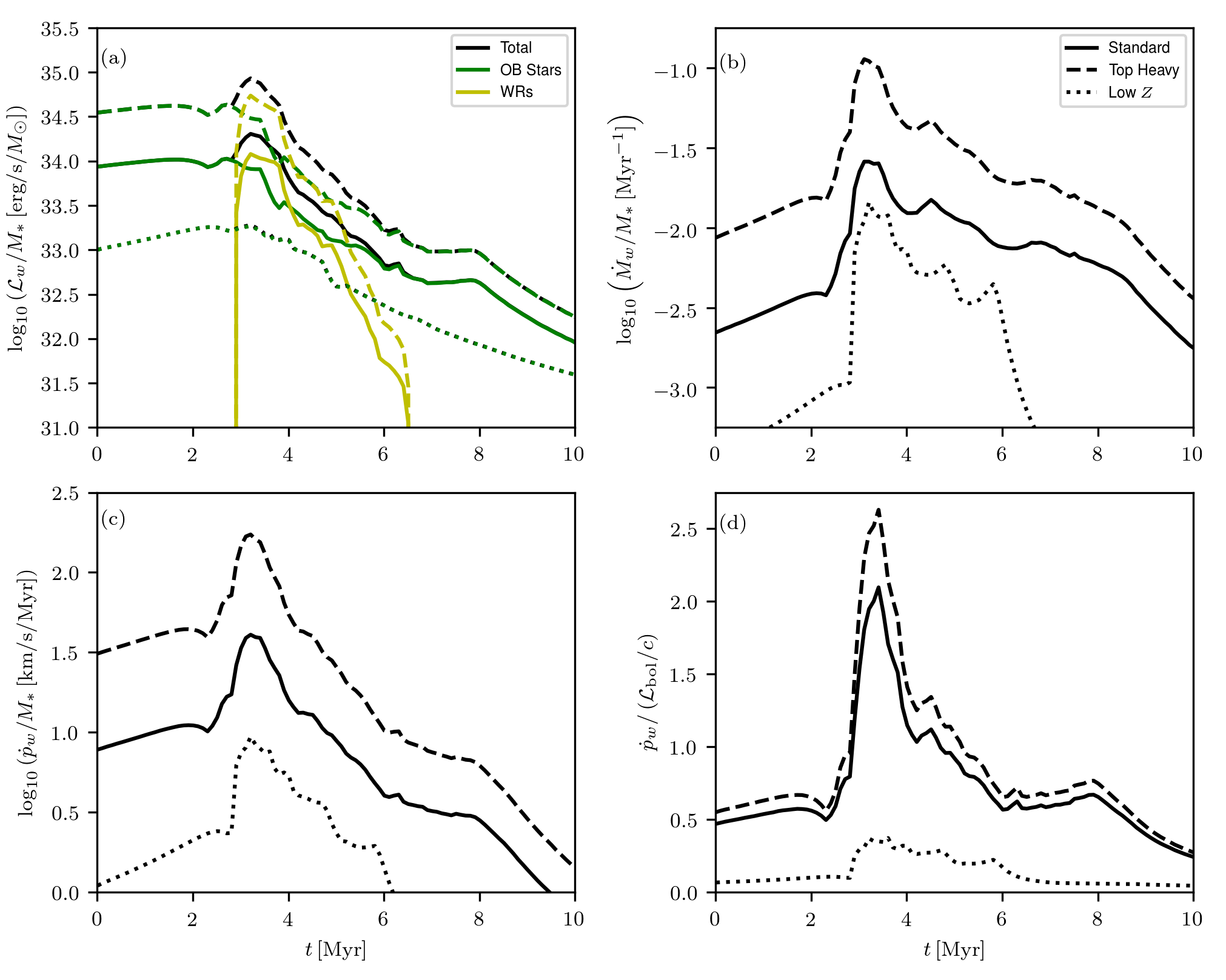}
    \caption{
    Evolution of several key wind quantities as calculated using the Starburst99 code \citep{SB99}. We show (a) the wind luminosity per unit mass of the star cluster, (b) mass-loss rate in the wind per unit stellar mass, (c) wind momentum injection rate per unit stellar mass, and (d) the ratio of wind momentum input rate to radiation momentum input rate. Solid lines show the result for a standard Kroupa IMF, dashed lines are for a ``top-heavy'' IMF in which $d\log N/d\log m$ is changed from $-2.3$ to $-1.8$ at $m>0.5\, M_{\odot}$, and dotted lines show the results for a standard 
    IMF but at low metallicity, $Z = Z_{\odot}/7$.
    }
    \label{fig:wind_evol}
\end{figure*}

\begin{figure*}
    \centering
    \includegraphics{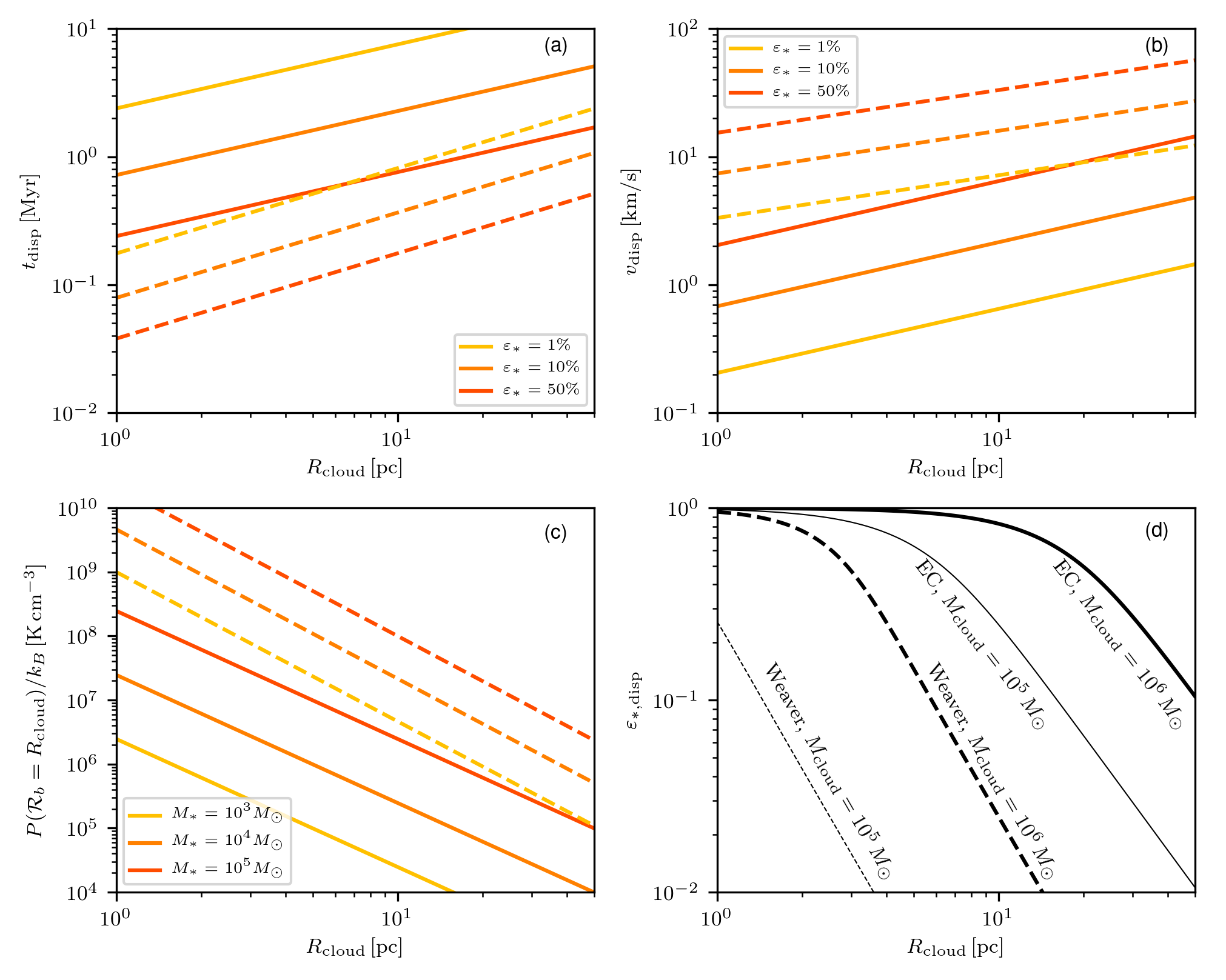}
    \caption{A comparison of several quantities as predicted by our theory 
    (solid lines) and that of \citet{Weaver77} (dashed lines). For the top 
    two panels we show results as a function of cloud radius and star formation 
    efficiency $\sfe=1,\, 10, \, 50\%$, while the bottom-left shows results 
    for $M_*=10^3, 10^4,\ 10^5 M_\odot$ (yellow, orange, and red respectively), 
    using $\Msh = 10^5 M_{\odot}$ for the dashed lines.. Panels show: {\it (a)} 
    shell velocity $v_\mathrm{disp}$ when the bubble has reached the edge of the 
    cloud; {\it (b)} time $t_\mathrm{disp}$ taken for this to occur; {\it  (c)} 
    the bubble interior pressure at this time. {\it  (d)} the star formation 
    efficiency needed to disperse the surrounding cloud. Here we show 
    $\varepsilon_{*,{\rm disp}}$ as a function of $\rcloud$ for 
    $\mcloud = 10^6,\, 10^5\, M_{\odot}$ displayed as the thick 
    and thin lines respectively.}
    \label{fig:weaver_comp}
\end{figure*}

\section{Application to Star-Forming Clouds}
\label{sec:applications}
Here we write down several formulae that will equip the reader to make practical 
use of our theory. To inform this discussion, we show in 
\autoref{fig:wind_evol} the evolution in time of several quantities related to 
stellar winds as calculated using the Starburst99 code \citep{SB99}. We show the 
evolution of the specific wind luminosity, the specific mass loss rate, the 
specific momentum input rate, and the so-called ``wind efficiency parameter,'' 
which is the ratio of momentum input rate in the wind and in radiation,   
$\pdot/( L_{\rm bol}/c)$.

For all quantities we show the evolution for both a standard solar metallicity 
($Z=0.014$), Kroupa initial mass function (IMF) \citep{KroupaIMF} (solid lines); 
for a ``top-heavy'' IMF with a shallower slope at $m>0.5M_{\odot}$ and solar 
metallicity (dashed lines); and for a standard IMF with 
$Z = 0.002\approx 10^{-0.85}Z_{\odot}$ (dotted lines, same as `low-$Z$' models 
in \citet{Leitherer14}). Additionally, in panel $a$ we separately show the 
contributions from OB stars (green lines), Wolf-Rayet (WR) stars (yellow lines), 
and the total (black lines). There are also very minor contributions to the total 
$\Lwind/M_*$ from luminous blue variable stars and red super-giant stars. We do 
not show these separately but they are included in the displayed total. The choice 
of the slope for the alternative IMF, $d\log N/d\log m=-1.8$ (instead of $-2.3$ 
for the standard IMF) is motivated by observations indicating that super star 
clusters, including those in the Galactic Center, may be top-heavy 
\citep{McCrady05,Lu13,Hosek19}. 

Our theory, as well as the simulations of Paper II, assumes that all the wind 
properties are constant in time. \autoref{fig:wind_evol} shows that this 
assumption is reasonable during the first $\sim 2\, {\rm Myr}$ of wind evolution, 
before the onset of significant energy input from WR stars.  For all of the 
numerical estimates below, we adopt the standard Kroupa IMF.  With a top-heavy 
IMF, $\dot p_w$ would be a factor of 4 higher, whereas $\dot p_w$ would be an 
order of magnitude lower with a low-metallicity IMF.

For the purposes of enabling quick calculations, the formulae we present below 
are written as simply as possible, setting $f_{\rm turb}=0$, $\alpha_R = 1$.  
Where appropriate, we comment on the range of  adjustments that would be 
expected based on the numerical results presented in Paper II.  
Additionally, in \autoref{fig:weaver_comp} we provide a comparison of the 
predictions of our theory and that of \citet{Weaver77} for several key 
quantities of interest.  As we shall show, the shell velocities, cloud 
dispersal timescales, and hot-gas pressures in the EC solution can be orders 
of magnitude below the corresponding values for the \citet{Weaver77} solution. 
This also implies much higher star formation efficiency would be required 
to disperse the cloud when interface cooling is taken into account.  

The primary application of our results is to bubbles collectively driven by 
the winds from young massive stars in clusters within star-forming clouds. We 
shall take the total stellar mass of the cluster (including both low- and 
high-mass stars) as $M_*$, and assume a centrally concentrated wind source. 
Using the result from SB99 with a standard Kroupa IMF, the direct wind momentum 
input rate is  
\begin{equation}
    \label{eq:SBmom} 
    \pdot = 8.6 \kms \Myr^{-1} M_*;
\end{equation}
here the coefficient is based on an average over the first Myr, while the 
value would be slightly lower ($\approx 7.85$) for an average over $0.1$ Myr. 
Averaging over the same time period, $\vw= 3512 \kms$.  Our solution allows for 
the momentum input rate to be enhanced by a factor $\alpha_p$ relative to the 
ideal ``momentum-conserving'' limit (see \autoref{eq:pEC}).

We shall assume that after a time $t$ the gas surrounding the central cluster 
has been swept up in a shell of mass $M_{\rm sh}$ by the wind. The shell's 
velocity will then be 
\begin{equation}\label{eq:shellvel}
    v_{\rm sh}= 8.6 \kms \alpha_p
    \frac{M_*}{M_{\rm sh}}
    \frac{t}{\Myr}.        
\end{equation}
When the input wind power is very low (e.g $M_*\lesssim 10^3 M_\odot$) we 
show in Paper II that $\alpha_p\sim 1.5-4$.
This range could apply in moderate density GMCs, where the efficiencies 
of star cluster formation are only $M_*/M_{\rm cloud} \equiv \sfe\sim 1-10\%$ 
\citep{Lada2010,Vutisalchavakul16}.  However, winds are likely to be most 
important relative to other feedback processes in very dense clouds, such 
as those forming super star clusters (SSCs), where $\sfe \gtrsim 10\%$ 
\citep{Leroy17,Emig20,Levy20}.  For this range, the momentum enhancement 
above the direct wind input is modest, $\alpha_p\sim 1.2- 2$ (see Paper II).

The evolutionary time is difficult to ascertain in observations, but 
from \autoref{eq:rbub_ecw} the shell velocity can also be expressed in 
terms of its mass $M_{\rm sh}$, radius $R_{\rm sh} = \reff$, and $\pdot$ as 
\begin{subequations}
\begin{eqnarray}
    v_{\rm sh} &=& \left(
    \frac{\alpha_p \pdot}{2} \frac{R_{\rm sh}}{ M_{\rm sh}}  \right)^{1/2} \\
      &=& 2.0 \kms \left( \alpha_p 
    \frac{M_*}{M_{\rm sh}} 
    \frac{R_{\rm sh}}{1\, \pc}  \right)^{1/2} \, 
    \label{eq:vshrsh}
\end{eqnarray}
\end{subequations}
where we have assumed $M_{\rm sh} = 4\pi R_{\rm sh}^3\bar\rho/3$. 
Given that $\alpha_p^{1/2} \sim 1-2$, \autoref{eq:vshrsh} makes 
clear that a  wind-driven shell can only reach velocity $> 10\kms$ 
if $M_* \gg  M_{\rm sh}$, assuming $R_{\rm sh}$ is in the range 
$\sim 1 - 25 \pc$ of observed star-forming clouds. We note, 
however, that a top-heavy IMF would increase the shell momentum 
and velocity somewhat.

We can also use \autoref{eq:rbub_ecw}, setting $\reff = R_{\rm cloud}$, 
to obtain an estimate for the time required for a wind-driven bubble 
from an embedded cluster to disperse the entire surrounding parent 
cloud.  For this simple estimate, we set $M_* = \sfe\mcloud$ and 
$M_{\rm sh} =(1-\sfe)\mcloud$.  The cloud dispersal time can then 
written as 
\begin{subequations}
\begin{eqnarray}
    \label{eq:tesc_est}
    t_{\rm disp} &=& \left( \frac{\ms}{2\alpha_p \pdot}
    \frac{1-\sfe}{\sfe }\rcloud \right)^{1/2} \\
    &=& 0.24 \Myr \left( \frac{1}{\alpha_p}
     \frac{1-\sfe}{\sfe }\frac{\rcloud}{1\, \pc} \right)^{1/2}.
\end{eqnarray}
\end{subequations}
The shell velocity $v_{\rm disp}$ at this time is obtained by replacing 
$M_*/M_{\rm sh} \rightarrow \sfe/(1-\sfe)$ and $R_{\rm sh} \rightarrow \rcloud$ 
in \autoref{eq:vshrsh}. It is interesting that these expressions are not 
(explicitly) dependent on the total cloud mass.  As before, the enhancement of 
the momentum input rate relative to the ideal ($\alpha_p=1$) EC solution would 
imply at most a $\sim 30\%$ reduction in $t_{\rm disp}$ for cases with 
$\sfe \gtrsim 10\%$.  A key point is that compact clouds with high 
$\sfe$ would be dispersed by wind feedback very rapidly, well before 
supernovae commence.

It is important to note that the expansion rate, and therefore other 
derived values such as the cloud dispersal time, are very different 
in the EC solution from what they would be if the classical 
\citet{Weaver77} solution applied.  For example, using the 
\citet{Weaver77} solution, equivalent calculations to the above 
would instead yield 
\begin{equation}
    \label{eq:tesc_w}
    t_{\rm disp, W} = 0.038\, {\rm Myr}
    \left(\frac{1-\sfe}{\sfe}\right)^{1/3} 
    \left(\frac{\rcloud}{1\,{\rm pc}}\right)^{2/3}\,
\end{equation}
and
\begin{equation}
 v_{\rm disp, W} = 15.4\, \kms
    \left(\frac{\sfe}{1-\sfe}\right)^{1/3} 
    \left(\frac{\rcloud}{1\,{\rm pc}}\right)^{1/3}\,    
\end{equation}
for the cloud dispersal time and the shell velocity at that time.

In the top two panels of  \autoref{fig:weaver_comp}, we compare the 
EC and Weaver solutions for the breakout time and corresponding velocity.
For the EC solution, the breakout velocity (breakout time) is a factor 
$\sim 5-10$ lower (higher).  Since \autoref{eq:tesc_w} underestimates the 
dispersal timescale by up to an order of magnitude, we caution against its 
use.  We also caution that the estimates in \citet{Weaver77} and 
\citet{maclow88} of the cooling time for wind-blown bubbles cannot be used 
to modify the solution of \autoref{eq:rbub_weaver}, as they are obtained 
under the assumption that the hot bubble interior is itself cooling, rather 
than cooling losses occurring due to turbulent mixing at the interface 
between hot and cool gas.  

Our above estimates, and the analysis and simulations of this paper 
and Paper II more generally, neglect the effects of gravity. 
In reality, the rate at which the expanding shell gains momentum will 
be instantaneously reduced by the gravitational force of the cluster on 
the shell, $G M_* M_{\rm sh}/R_{\rm sh}^2$, and the gravitational force 
of the shell on itself, $G M_{\rm sh}^2/(2 R_{\rm sh}^2)$.  Allowing for 
gravity, the wind can drive expansion only provided
\begin{equation*}
    M_{\rm sh} + \frac{M_{\rm sh}^2}{2 M_*}
    < \alpha_p \frac{\pdot}{M_*} \frac{R_{\rm sh}^2}{G} \, .
\end{equation*}

For a self-gravitating cloud to be fully dispersed by an expanding 
wind bubble, we again set $R_{\rm sh}= \rcloud$,
$M_* = \sfe\mcloud$, and $M_{\rm sh} =(1-\sfe)\mcloud$, and solve 
the inequality to obtain a condition on the SFE, $\sfe$.  We express 
this condition in terms of the original surface density of gas in the 
cloud, $\Sigma_{\rm cloud} \equiv  \mcloud/(\pi \rcloud^2)$:
\begin{subequations}
\begin{eqnarray}\label{eq:SFEgen}
    \frac{1-\sfe^2}{\sfe} &<& 
    \frac{2\alpha_p}{\pi}\frac{\pdot}{M_*}\frac{1}{ G  \Sigma_{\rm cloud}}\\
    &=& 1.2 \alpha_p 
    \left(\frac{\Sigma_{\rm cloud}}{10^3 \Msun \pc^{-2}}  \right)^{-1} \, .\label{eq:SFEwind}
\end{eqnarray}
\end{subequations}
The point at which $\sfe$ is large enough that this inequality is 
satisfied is a simple estimate for the star formation efficiency needed 
to disperse the cloud, $\varepsilon_{*,{\rm disp}}$. For ``normal'' GMCs 
with $\Sigma_{\rm cloud}\sim 10^2  \Msun \pc^{-2}$, 
\autoref{eq:SFEwind} reduces to 
\begin{equation}\label{eq:SFElow}
    \varepsilon_{*,{\rm disp}} \approx \frac{0.08}{\alpha_p}
    \frac{\Sigma_{\rm cloud}}{10^2 \Msun \pc^{-2}} \, .
\end{equation}
For high surface density clouds 
$\Sigma_{\rm cloud}\sim 10^4 -10^5 \Msun \pc^{-2}$ such as those 
forming SSCs, this reduces to 
\begin{equation}\label{eq:SFEhigh}
    1-\varepsilon_{*,{\rm disp}} \approx 0.06 \alpha_p 
    \left(\frac{\Sigma_{\rm cloud}}{10^4 \Msun \pc^{-2}}  \right)^{-1}\, .
\end{equation}
Clearly, extremely high surface densities would translate to an expected SFE 
very close to unity if winds were the only form of feedback acting.  In the 
bottom right panel of \autoref{fig:weaver_comp}, we compare the estimate for 
$\varepsilon_{*,{\rm disp}}$ based on the EC solution with an equivalent 
calculation based on the \citet{Weaver77} theory.  Evidently, the loss of 
energy to  cooling renders winds far less effective at limiting star 
formation in molecular clouds.  

It is worth noting that \autoref{eq:SFEgen} also applies in estimating the SFE 
for momentum sources other than winds, by substituting an alternative specific 
momentum input rate for  $\alpha_p \pdot/M_*$.  For example, for the same 
fully-sampled Kroupa IMF \citep{KroupaIMF}, SB99 gives a specific momentum 
injection rate from radiation of $L_*/(c M_*) \approx 20 \kms \, \Myr^{-1}$, a 
factor of 2.3 larger than that stellar wind momentum input rate in 
\autoref{eq:SBmom} (see \autoref{fig:wind_evol}).  With $\alpha_p \sim 1-4$ as 
obtained from our simulations (Paper II), the direct momentum input rates from 
radiation and winds would be comparable. Allowing for the combined momentum 
injection rate would respectively reduce the RHS of \autoref{eq:SFElow}, or 
increase the RHS of \autoref{eq:SFEhigh}, by a factor of $\sim 2$.  

The above simple estimates of SFEs are interesting, but important caveats should 
be kept in mind.  First, there is an implicit assumption that all wind momentum 
is retained in a cloud, but in fact it is likely that a significant portion is 
lost once the bubble size becomes comparable to that of the parent cloud, because 
the fractal bubble can break out through fingers and vent the shocked wind gas.  
For radiation feedback, estimates of the SFE based on the simple spherical 
assumption significantly underestimate the SFE obtained from simulations with 
realistic turbulent clouds due to cancellation from multiple sources and escape 
through low-density channels 
\citep[e.g.,][]{dale17,Raskutti16, raskutti17,JGK18,kim19}, and this is likely to 
be true for winds as well.  
Thus, \autoref{eq:SFElow} and \autoref{eq:SFEhigh} should not be taken too 
seriously as direct predictions but more properly as bracketing the potential 
impact of winds on $\sfe$.   

Second, the effects of mass loss due to photoevaporation 
and ionized gas pressure acting on neutrals are not considered above. In fact, at 
low and moderate surface densities photoionization (and to a lesser extent direct 
radiation pressure) is quite effective in limiting star formation and dispersing clouds 
\citep[e.g.][]{Dale12,KrumholzMatzner09,Raskutti16,Geen17,JGK18,Grudic18,he19,JGK20}.
Our expectation is that stellar winds would aid in quenching star formation, but are 
likely to rival the importance of photoevaporation only in clouds of quite 
high surface density (see \autoref{sec:discussion}).  

Our results also provide an estimate for the thermal pressure of hot 
gas in the interior of the bubble, which is a key observable. 
\autoref{eq:pressure_time} gives the pressure as a function of bubble 
radius $\reff$; writing this in dimensional form,
the result is 
\begin{equation}
    \label{eq:pressure}
    \frac{\Pbub}{k_B} \approx 2.46
    \times 10^6 \,{\rm K}\, {\rm cm}^{-3}  
    \frac{M_*}{10^3 \Msun} 
    \left(\frac{\reff}{1\, \pc}\right)^{-2}\, \frac{3 \alpha_p -1}{2}.
\end{equation}

In the bottom left panel of \autoref{fig:weaver_comp}, we compare the bubble 
pressure when $\reff = \rcloud$ to the bubble pressure predicted by the 
\citet{Weaver77} theory for several different cluster masses. It is 
striking that the pressures differ by two or three orders of magnitude over 
much of the parameter space.

Even with cooling losses, the pressure given by \autoref{eq:pressure} would be 
quite high for small bubbles powered by luminous, massive clusters.  Larger 
bubbles around lower-mass clusters would have much more moderate pressure.  For 
the wind bubbles around single O stars rather than stellar clusters, the predicted 
pressure in hot gas is
\begin{align}\label{eq:pressure_single}
    \frac{\Pbub}{k_B} \approx & 2.86 \times 10^5 \,{\rm K}\, {\rm cm}^{-3}  
    \nonumber\\
    &\times\frac{\dot p_{w,1}}{
    10^3\Msun \kms \Myr^{-1}} 
    \left(\frac{\reff}{1\, \pc}\right)^{-2} \frac{3 \alpha_p -1}{2},      
\end{align}
where $\dot p_{w,1}=(2 \dot E_{w,1} \dot M_{w,1})^{1/2}$ is the single-star 
wind momentum input rate. 

For our simulations, as presented in Paper II (see Figs. 1 and 2 there), 
the hot gas pressure is far lower than that predicted in the original Weaver 
solution, because most of the energy deposited by the wind is lost to cooling. 
Paper II shows that the measured energy reduction factor 
$1-\Theta \sim 0.1 -0.01$ is in good agreement with \autoref{eq:Theta_cool_prediction}. 
Taking $\ratr \approx \alpha_p$, the instantaneous fraction of the injected 
wind energy that is not radiated away is predicted to be 
\begin{equation}
    1- \Theta \approx 4\times 10^{-3}\alpha_p \frac{v_{\rm sh}}{10 \kms}  
\end{equation}
in terms of the shell's expansion velocity $v_{\rm sh} = \dot{\cal R}_b$. 
Even if this velocity is not measured directly, it may be estimated from other 
observables through \autoref{eq:vshrsh}.  

The comparisons of this section between results obtained using the EC solution 
and the classical \citet{Weaver77} solution show that neglect of energy losses 
due to interface mixing can lead to erroneous conclusions regarding the importance 
of stellar winds compared to other feedback mechanisms.  For star clusters prior 
to breakout from molecular clouds, the EC conditions of this paper will generally 
apply, and we recommend use of formulae in this section.  For a cluster after 
breakout or for runaway stars, the ambient density would be lower, leading to a 
longer cooling time $\tcool$.  Lower power sources than the clusters of 
$M_* \gtrsim 10^3 \Msun$  considered here would produce lower turbulent velocities 
$\vt$. In the longer-$\tcool$, smaller-$\vt$ situations where interface energy 
losses are more moderate, we instead recommend use of the formulae in 
\citet{ElBadry19} with an appropriate diffusion parameter (such as 
\autoref{eq:kappa_corr}).

\section{Discussion}
\label{sec:discussion}

\subsection{Observations}
\label{subsec:obs_discuss}

This work was inspired in part by observations of evolved HII regions, 
which suggest that the classical model of \citet{Castor75} and 
\citet{Weaver77} -- in which nearly half of the wind energy remains in a 
hot bubble that can emit X-rays and drive rapid shell expansion 
($\sim 10-50\kms$ from \autoref{fig:weaver_comp}) -- is not 
in agreement with observations   
\citep{Townsley03,Townsley06,HCM09,Lopez11,Lopez14,Rosen14}.
Instead, observations suggests that wind energy must be lost either through 
leakage \citep{HCM09} or through radiative cooling  at intermediate 
temperatures.  The former is encouraged by the highly inhomogeneous structure
of clouds, while the latter is facilitated by turbulent mixing. We have 
explored the consequences of turbulent mixing and cooling as a major energy 
sink for shocked wind gas. In Paper II, we show that such extreme cooling 
is not only possible but quite typical. These effects are expected to be 
especially important for early evolution before wind bubbles break out 
of their natal cloud.  We have proposed (and show in Paper II) that the 
interfaces between hot and cool gas in wind bubbles are fractals; this is 
crucial in enhancing mixing, and also important to breakout as bubble 
``fingers'' can vent gas earlier than would otherwise occur. 

The EC model may be able to explain several observations.
\citet{Rosen14} showed that, for the four dense star clusters they studied, 
only 3-30\% of the energy deposited by stellar winds could be accounted for 
by radiative cooling internal to the bubble (negligible for the most part) 
combined with mechanical work on the surroundings (Bottom left panels of 
Figures 6-9 in that work), suggesting up to 97\% of the wind energy is lost. 
While \citet{Rosen14} were able to explain the missing energy in some of the 
clusters by appealing to thermal conduction (under the assumption that 
energy is lost, rather than leading to evaporation) and dust-reprocessed 
radiation, these calculations are more uncertain and sometimes over-account 
for the lost energy.  \citet{Rosen14} also raised the possibility of significant 
energy losses by ``turbulent conduction'' (turbulent mixing+cooling in our 
terminology), but they did not attempt to estimate this.  Since we find (see 
Paper II, Figure 17) that as little as $1\%$ of the input  energy remains in 
our simulation, we conclude that radiative cooling in turbulent mixing layers 
would easily account for the missing wind energy in observed star-forming regions.

\citet{Olivier20} recently observed deeply embedded HII regions, where our 
theory should be most applicable, but were only able to put upper limits on 
effects of stellar winds due to a lack of long-exposure X-ray observations. 
Further work on this front should allow a quantitative test of our theory.

The nearby Orion Molecular Cloud has some of the best observations 
of on-going star formation feedback in action. Using X-ray observations 
with the XMM-Newton satellite, \citet{Gudel08} showed that the bubble 
being driven by the Trapezium cluster is pervaded by 
$1.7 - 2.1\times 10^6 \, {\rm K}$ gas, and estimated an electron density 
$n_e = 0.2-0.5 \, {\rm cm}^{-3}$, implying thermal pressure of 
$\Pbub/k_B= 2 n_e T = 0.68-2.1\times 10^6 {\rm K}\, {\rm cm}^{-3}$.
Using the wind parameters quoted for $\theta^1$ Ori C (the most massive 
star in the Trapezium) by \citet{Gudel08}, 
$\mdot = 0.8 \, M_{\odot}/{\rm Myr}$ and $\vw = 1650 \, \kms$, we get 
$\pdot = 1.32\times 10^3\, M_{\odot}\, {\rm km/s/Myr}$. Using 
\autoref{eq:pressure_single} and an effective radius 2 pc (\citealt{Gudel08} 
suggest a range 1 - 4 pc, uncertain due to geometry) we find we would require 
$\alpha_p \approx 5$ to obtain pressure agreement at the lower end of the 
estimated range. This value of $\alpha_p$ is similar to the results we find 
in our simulations for our lowest momentum input rate, 
$\pdot = 1.8 \times 10^4 \, M_{\odot}\, {\rm km/s/Myr}$ (Paper II), considering 
the considerable uncertainties in the observational parameters; for example, 
substituting estimated wind momentum input rate of \citet{Gagne05} would 
increase the pressure by a factor of two.

Our work helps to explain why observational (listed above) and numerical 
\citep{Dale14,Geen15b,Geen20} works have found winds to be inefficient at 
cloud dispersal compared to effects from ionizing  radiation. However, taking 
these results to mean that winds are unimportant is an over-simplification. 
Winds remain important not only because they pollute the gas surrounding 
nascent star clusters, perhaps changing the chemical composition of 
subsequent stellar populations \citep[e.g.][]{BastianLardo18,Gratton19}, 
but also because they may be the primary mechanism for cloud dispersal 
{\textit{in some environments}}.

Very recently, new empirical evidence of feedback in extreme environments was 
obtained by \citet{Levy20}, who observed the environment of nascent super star 
clusters in NGC 253 with $0.5 \, {\rm pc}$ resolution ALMA observations in 
multiple molecular lines. From analysis of P-Cygni profiles, indicative of 
cluster-scale outflows, \citet{Levy20} obtained estimates of the mass and 
momentum carried by outflows around three of the SSCs in NGC 253 and applied 
our theory of stellar wind feedback to test whether this could explain the 
observations.  They found that given modest momentum enhancement factors, 
$\alpha_p\sim 1-4$, winds could plausibly be the dominant feedback mechanism. 
However, our numerical findings (see Paper II Figure 8) suggest that 
$\alpha_p\sim 1-1.5$ in the dense environments of SSCs (assuming high SFE).
Given the uncertainty in the measurements, winds could still be the main 
drivers of feedback in these systems. The inferred $\alpha_p \gtrsim 2$ 
could, however, also be a sign of the contribution of other feedback 
mechanisms, such as radiation. A top-heavy IMF (see \autoref{fig:wind_evol}) 
or a super-solar metallicity (not shown) stellar population could also boost 
the input momentum/stellar mass \citep{Leitherer92,Vink01,Leitherer14}.

\subsection{Other models of momentum-driven bubbles}
\label{subsec:models_dicuss}

In Section~\ref{sec:applications}, we explicitly compared our results to what 
would be predicted based on the classic ``Weaver-type'' solution described in
Section~\ref{subsec:classic_wind}.  This type of solution can be classified as
``energy-driven'' evolution because the energy of the bubble increases linearly 
in time, with the bubble radius expanding as $\reff \propto t^{3/5}$. In 
contrast, in our EC solution it is the \textit{momentum} that increases linearly 
in time, while energy only increases as the square-root of time.  Evolution in 
the EC solution may therefore be classified as ``momentum-driven.'' 

Momentum-driven solutions for wind bubble expansion are characterized by 
$\reff \propto t^{1/2}$, and have been considered by several previous authors.
\citet{Steigman75} simply assumed momentum-driven evolution, disregarding the 
importance of shocks and internal bubble structure.  \citet{Avedisova72}, 
\citet{Castor75}, and \citet{Weaver77} clarified the role of internal wind 
shocks and argued for the importance of energy build-up within the bubble. 
Subsequent authors discussed the possibility of momentum-driven evolution due 
to enhanced cooling within the bubble interior, due to a variety of mechanisms 
described below. Two key differences of our theory from these earlier proposals 
are that (1) in our model energy is lost through strong radiative cooling at 
intermediate temperature in a turbulent mixing layer at the surface of the 
bubble, and (2) fractal structure of the interface is essential to efficient 
cooling.    

The wind models of \citet{KooMcKee92a,KooMcKee92b} specify 
a so-called ``slow wind'' which leads to momentum-driven bubble evolution. 
However, in this model the momentum-driven evolution is caused by 
the interior of the bubble being radiative when the wind is sufficiently 
dense and slow, where ``slow'' here is hundreds of $\kms$, much slower 
than the $\vw >10^3 \, \kms$ winds under consideration in this paper.

Similarly, Section 7B of \citet{OstrikerMcKee88} considers that interior 
radiative losses could be enhanced due to the evaporation of clouds 
overtaken by the wind \citep[c.f.][]{CowieMcKee77}. In this scenario,
the majority of the cooling would still occur in the shocked bubble 
interior.

In \citet{SilichTT13} as well as \citet{maclow88} (following 
\citealt{Weaver77}) it is emphasized that conduction acting at the 
boundary between the shocked wind and swept-up material creates a 
temperature gradient and includes a regime where cooling would be 
efficient, which would eventually lead to the evolution becoming  
momentum-driven. This is the scenario most similar to our model. However, 
as discussed in \citet{SilichTT13}, for typical parameters this transition 
is only expected to occur after $\sim 10 \, {\rm Myr}$, long past the 
time-scales of evolution considered here or over which the wind can be 
treated as constant-luminosity (as evidenced by \autoref{fig:wind_evol}). 
In our model, radiative cooling in a turbulent mixing layer is far 
more efficient than the conduction-induced cooling, due in part to the fractal 
nature of the interface, and the bubble becomes momentum-driven at much 
earlier times.

\subsection{Comparison to Radiative Feedback}
\label{subsec:radiation_disucss}

Several studies have argued analytically or demonstrated numerically that 
radiation feedback becomes much less efficient at dispersing clouds and 
limiting star formation in very high density environments 
\citep[e.g.][]{Fall10,Dale12,JGK16,Raskutti16,JGK18,Grudic18,Rahner19,Fukushima20}, 
even allowing for reprocessed radiation  \citep{Skinner15,Tsang18}.  In such 
dense clouds, the evolutionary timescales are also too short for supernovae 
to be important in dispersing cloud material. 

Quantitatively, \citet{JGK18} (Figures 8 and 12) show that when the surface 
density of a cloud exceeds $\sim 10^3 \Msun \pc^{-2}$, the total momentum 
injection rate per stellar mass due to EUV and FUV radiation pressure\footnote{From 
Fig. 7 of \citet{JGK18}, direct radiation pressure dominates over effects of 
photoevaporation and ionized gas pressure in this regime.} drops below that in 
\autoref{eq:SBmom} for winds. This suggests that winds could potentially be the 
dominant feedback process in higher density environments. In particular, our theory 
may be able to explain several aspects of the extremely compact molecular clouds 
seen to be forming super star clusters in nearby galaxies 
\citep{Johnson15,Oey17,Turner17,Leroy18,Emig20}. As super star clusters/young 
massive clusters \citep[e.g.][]{Whitmore03,Portegies-Zwart10} are the densest 
redshift-zero loci of star formation, understanding their formation is not 
only interesting in itself, but provides a window on globular cluster 
formation.

It may be noted that the linear dependence of $\sfe$ on surface density 
given by \autoref{eq:SFElow} is similar to the case of radiation-pressure 
driven cloud dispersal under the idealization of a spherical system 
(\citealt{Fall10,JGK16}; see also \citealt{Grudic18,Li19}). This differs from 
the $\sfe \propto \Sigma_{\rm cloud}^{1/2}$ dependence reported by 
\citet{Fukushima20}. A relation $\sfe \propto \Sigma_{\rm cloud}^{1/2}$ is 
close to what would be predicted if the cloud lifetime scales linearly with 
the propagation time of a self-similarly expanding ionization front, while 
the star formation rate scales inversely with the cloud free-fall time.  
However, the generality of the \citet{Fukushima20} result may be questioned, 
since their simulations adopted very strongly self-gravitating clouds with a 
limited range of initial gas surface density.  In contrast, the 
radiation-hydrodynamic simulations of \citet{JGK18} found a non-power-law 
scaling of $\sfe$ with $\Sigma_{\rm cloud}$ (see their Eq. 26) over a wider 
parameter range for clouds that were initially marginally bound. The difference 
between simple scaling predictions and full numerical results for the case of 
radiation-driven cloud destruction serves as a caution against taking 
\autoref{eq:SFEgen} as more than a general indicator of the trend in SFE.  
Full numerical simulations will be needed to assess the effectiveness 
of winds in limiting star formation in clouds of varying conditions.

Regardless of the environment, winds may be more important than ionizing 
radiation at early stages of evolution when the radius of the \ion{H}{2} 
region is small. Similar to the approach in \citet{KrumholzMatzner09},
one can make a simple comparison of the wind momentum input rate 
$\alpha_p \pdot$ to the nominal force imposed by thermal pressure of 
photoionized gas.  For an effective bubble radius $\reff$, the latter is 
$2n_e k_{\rm B} T_{\rm II}(4\pi \reff^2) \propto (Q_{\rm i}\mathcal{R}_b)^{1/2}$, 
where we have used
$n_e=[3 f_{\rm ion} Q_{\rm i} /(\alpha_{\rm B} 4 \pi \reff^3 )]^{1/2}$ 
for ionization equilibrium with a source of ionizing photon rate $Q_i$ 
and a fraction $f_{\rm ion}$ of photons available for ionization, and 
$T_{\rm II}$ denotes the temperature of the photoionized gas. The 
force from the wind would exceed that from radiation when the radius 
is smaller than a characteristic radius
\begin{equation}
    \mathcal{R}_{b,{\rm ch}}  = \frac{\alpha_{\rm B}}{12 \pi (2k_{\rm B}T_{\rm II})^2}\frac{\alpha_p^2 \pdot^2}{f_{\rm ion}Q_{\rm i}}\,. 
\end{equation}
Adopting $T_{\rm II} = 8000 \Kel$, 
$\alpha_{\rm B} = 3.03 \times 10^{-13}\,{\rm cm}^3\,{\rm s}^{-1}$, 
and $f_{\rm ion}=0.5$, the characteristic radius for a bubble driven by a 
single massive star with $\pdot = 10^4 \Msun\kms\Myr^{-1}$ and 
$Q_{\rm i} = 4\times 10^{49}\,{\rm s}^{-1}$ is 
$\mathcal{R}_{b,{\rm ch}} = 0.11 \alpha_p^2\, \pc$. For a star cluster 
with $Q_{\rm i} \approx 4\times 10^{46}\,{\rm s}^{-1} (M_*/\Msun)$ and 
$\pdot = 8.6 \kms\Myr^{-1} M_*$, we obtain 
$\mathcal{R}_{b,{\rm ch}} \approx 0.79 \alpha_p^2 (M_*/10^4 \Msun) \pc$; 
the wind is expected to play a greater role than photoionization for 
bubbles created by massive cluster stars and in early stage of evolution.

Therefore, similar to the situation analyzed by \citet{JGK16} where the 
role of wind momentum input is played by radiation pressure, for 
sufficiently small radii it would be expected that the wind dominates the 
dynamics.  The ionized gas layer would be compressed by the wind into a narrow 
layer near the ionization front. The wind would indirectly drive the shell 
expansion by enhancing density and thermal pressure of the \ion{H}{2} region. 
The relative volumes of the hot-gas and photoionized-gas regions can be obtained 
by requiring that the pressures match.  In the regime where winds dominate, 
radiation pressure is expected to contribute to the expansion by further 
increasing the density at the ionization front \citep{JGK16,Rahner17}.
Later in evolution when the radius of the \ion{H}{2} region is large, the 
photoionized gas would instead compress the hot interior of the bubble.  
Expansion would be driven primarily by photoionized gas that fills most 
of the bubble volume.

We note that the quantitative predictions above are based on the assumption 
that there is no significant leakage of hot gas and radiation. At late stage 
of bubble evolution ($\reff \sim R_{\rm cloud}$), both (hot and photoionized) 
gas and radiation are expected to break out through low-density holes created 
by turbulence.

\subsection{Other Physics}
\label{subsec:physics_discuss}

The theory presented here, as well as the simulations described in Paper II, 
are highly idealized, not considering many aspects of physical processes and 
the astronomical parameter space that could be important to the problem at 
hand. We briefly discuss some of these issues here.

First, the relevance of our results may be metallicity dependent. We note 
that wind mass loss is almost linear with metallicity, 
$d\log \dot{M}_w/d\log Z {\sim 0.7}$, while the wind velocity has a weak 
dependence on metallicity  ($V_w \propto Z^{0.13}$), so that the wind 
momentum input rate is almost linear in metallicity \citep{Vink01}. Lower 
metallicity in the cloud could in principle also make cooling less 
effective.  However, we show in Paper II that most cooling in the 
turbulently-mixed interface gas occurs near $T\sim 10^4 \, {\rm K}$ 
where cooling is mostly due to Hydrogen (Lyman $\alpha$), implying a 
change in metallicity should not have a strong effect.  Therefore, the 
effect of winds is expected to decrease at lower metallicity.

Second, we ignore thermal conduction. As noted above and in 
\citet{ElBadry19}, thermal conduction can act to increase the density 
and decrease the temperature of the bubble interior at fixed pressure. 
While it is not known whether the effect of thermal conduction in the 
presence of genuine turbulent mixing acts the same as for a parameterized 
diffusivity, if this is the case we might still expect the thermal 
pressure to follow Equations~\ref{eq:pressure} and \ref{eq:pressure_single}. 

Finally, while we do not include magnetic fields in our models, in general 
we would expect them to have at least two effects. Firstly, they would add 
to the external pressure and tension that the bubble must ``fight against'' 
to expand. However, our power-law solutions already neglect external 
turbulent and thermal stresses; when accounted for, these and the magnetic 
stresses would slow expansion when $\dot p_w/(4 \pi \reff^2)$ becomes 
comparable to the total external stress (and depending on the field geometry 
the bubble may also expand preferentially along the field lines). Secondly, 
a magnetic field could act to inhibit development of the turbulent mixing 
that leads to cooling.  For example, the Kelvin-Helmholtz instability is 
suppressed if the Alfv\'en speed in the shell gas exceeds 
$v_{\rm shear}/\chi^{1/2}$, where $v_{\rm shear}$ is the shear velocity and 
$\chi$ is the ratio between the density of shell gas and shocked wind gas.  
Even if primary instabilities take place, magnetic tension may limit 
turbulent cascades to small scales. Also, inclusion of radiation, as 
discussed above, would imply that the hot gas may principally interact with 
a layer of photoevaporated gas rather than directly with the shell  gas.  
Numerical study will be needed to assess the parameter regime where the 
mixing/cooling process and bubble evolution are strongly affected.

\section{Conclusion}
\label{sec:conclusion}

We have presented a theory for stellar wind bubble evolution with extremely 
efficient energy losses to radiative cooling, such that bubble expansion is 
determined by the original wind momentum injection rate.  In this theoretical 
model, the extreme energy losses are the result of turbulent mixing and 
subsequent radiative cooling at the interface between hot and cool gas at the 
bubble surface, strongly enhanced by the large interface area that arises from 
the fractal geometry of the bubble.

In a companion paper, referred to throughout this work as Paper II, we validate  
our theory using three-dimensional hydrodynamic simulations of wind-driven 
bubble evolution, demonstrating excellent agreement. In Paper II we quantify the 
few free parameters of our theory and provide a full analysis of thermal and 
dynamical structure and evolution.

The treatment of stellar winds in this paper and its companion has been 
largely simplified (constant luminosity with central point-like 
star clusters put in by hand). This treatment allowed us to gain a deeper 
theoretical understanding of the stellar wind bubble physics that controls 
evolution.  Still, a fully self-consistent treatment of star formation 
with wind feedback for a range of cloud parameters -- both normal GMCs 
and clouds capable of forming SSCs -- remains an important goal.  The 
fractal theory developed here will serve as a valuable tool aiding 
interpretation in this kind of numerical effort, as well as observational 
studies of star-forming cloud structure and evolution. In addition, the 
theory developed here may potentially be applied to understanding the 
history of chemical enrichment of natal clouds by embedded clusters, 
which would illuminate many longstanding problems involving multiple 
stellar populations within globular clusters 
\citep[e.g.][]{Wunsch17,LochhaasThompson17,BastianLardo18,Gratton19}.

\acknowledgments

We thank Drummond Fielding and Erin Kado-Fong for useful discussions. 
We thank the referee for their many insightful comments that improved 
the quality of the manuscript. We thank Cameron Lancaster for the 
production of the schematics shown in \autoref{fig:schematics}.
This work was partly supported by the National Science Foundation (AARG award
AST-1713949) and NASA (ATP grant No. NNX17AG26G). J.-G.K. acknowledges support 
from the Lyman Spitzer, Jr. Postdoctoral Fellowship at Princeton University. 
Computational resources were provided by the Princeton Institute for 
Computational Science and Engineering (PICSciE) and the Office of Information 
Technology’s High Performance Computing Center at Princeton University.

\software{
{\tt scipy} \citep{scipy},
{\tt numpy} \citep{harrisNumpy2020}, 
{\tt matplotlib} \citep{matplotlib_hunter07},
{\tt adstex} (\url{https://github.com/yymao/adstex})
}

\appendix

\section{The Wind Interior}
\label{app:wind_interior}

Here we provide a derivation for some of the properties of the free 
and shocked wind phases. This analysis allows us to write the energy 
enhancement factor $\ratr$, post-shock pressure, and the relative 
volume of free and shocked wind in terms of a ``momentum enhancement 
factor'' $\alpha_p$. 

We assume spherical geometry for the bubble with $r<\rfree$ occupied 
by the free wind phase, where 
$\rfree = (3V_{\rm f}/4\pi)^{1/3}$ for $V_{\rm f}$ the volume occupied 
by the free wind, and $\rfree< r< \reff$ occupied by shocked wind gas. 

In the free wind the flow expands with hypersonic wind velocity $\vw$ 
and the thermal energy is negligible, so that 
\begin{equation}
    \label{eq:free_wind}
    \rho (r) = \frac{\mdot}{4\pi r^2 \vw} \; \; , \; \; v_r = \vw \, .
\end{equation}
At $\rfree$ the wind then goes through a strong shock (assumed 
perpendicular), with immediate post-shock conditions
\begin{equation}
    \label{eq:rhops}
    \rhops = 4 \rho(\rfree) = \frac{\mdot}{\pi \rfree^2 \vw } \, ,
\end{equation}
\begin{equation}
    \label{eq:vps}
    \vsw = \frac{\vw}{4}   \, ,
\end{equation}
and
\begin{equation}
    \label{eq:Pps}
    \Pps = 3\rhops \vsw^2 = \frac{3\pdot}{16\pi \rfree^2 }   \, .
\end{equation}
The shocked wind region is assumed to have a steady flow,
\begin{equation}
    \label{eq:masscons}
      \frac{\rho}{\rhops}   =  \frac{\rfree^2}{r^2} \frac{\vsw}{v} \, ,
\end{equation}
with an adiabatic equation of state
\begin{equation}
    \label{eq:adiabat}
    \frac{P}{\rho}= \frac{\Pps}{\rhops}\left(\frac{\rho}{\rhops} \right)^{2/3} \, ,
\end{equation}
and Bernoulli parameter
\begin{equation}
    \label{eq:bernoulli}
    \frac{1}{2}v^2 + \frac{5}{2}\frac{P}{\rho} = 
    \frac{1}{2}\vsw^2 + \frac{5}{2}\frac{\Pps}{\rhops}\, .
\end{equation}
By substituting in for $P$ and $\rho$ in the Bernoulli equation 
we obtain
\begin{equation}
    \left(\frac{v}{\vsw} \right)^2  + 15 \left( \frac{\vsw}{v} 
    \left( \frac{\rfree}{r}\right)^2 \right)^{2/3} = 16, 
\end{equation}
which can be rearranged to arrive at 
\begin{equation}
    \label{eq:usolv}
    \frac{v}{\vsw} = \left( \frac{\rfree}{r} \right)^2 \left( \frac{15}{16 - (v/\vsw)^2}\right)^{3/2}\, .
\end{equation}
While this equation does have an analytic solution, it is quite 
complicated; an approximation that is good to within 10\% is
\begin{equation}
    \label{eq:u_approx}
     \frac{v}{\vsw} \approx  \left( \frac{\rfree}{r}\right)^2 \, .
\end{equation}
With \autoref{eq:u_approx}, both the density and pressure are approximately constant 
throughout the post-shock region, equal to their immediate post-shock 
values  given in \autoref{eq:rhops} and \autoref{eq:Pps}, $\rho=\rhops$ 
and $P=\Pps$ 

The total rate of momentum transport across a surface of radius $r$ is  
\begin{equation*}
    \dot{p}(r) = 4\pi r^2 \left(P + \rho v^2 \right)\, ,
\end{equation*}
which in the free wind region is $\dot p(r)=\pdot$. Using our solution 
in the shocked-wind region, this becomes 
\begin{equation}
    \dot{p}(r) = \frac{\pdot}{4} \left[3\left(\frac{r}{\rfree}\right)^{2}+\left(\frac{r}{\rfree}\right)^{-2} 
    \right] \, .
\end{equation}
This function is strictly increasing for $r/\rfree>1$.

The total momentum input to the shell is obtained by taking $r \to \reff$, 
which from the definition in \autoref{eq:momentum_ecw} implies
\begin{equation}
    \label{eq:alphap_derive}
    \alpha_p = \frac{1}{4}\left[3\left(\frac{\reff}{\rfree}\right)^2 + \left(\frac{\reff}{\rfree}\right)^{-2}  \right] \, ,
\end{equation}
 We can invert to obtain $\reff/\rfree$ in terms of $\alpha_p$ as 
\begin{equation}
    \label{eq:RbRfsoln}
    \left(\frac{\reff}{\rfree}\right)^2 = \frac{2}{3}\alpha_p + \left[ \left(\frac{2}{3}\alpha_p\right)^2 - \frac{1}{3}\right]^{1/2}.    
\end{equation}
A useful and simple approximation for the above, which is good to within 
4\% for $1\leq \alpha_p \leq 4$ is given by
\begin{equation}
    \label{eq:RbRf_approx}
    \left(\frac{\reff}{\rfree}\right)^2 \approx \frac{1}{2}\left(
    3\alpha_p - 1\right)\,;    
\end{equation}
alternatively, at large $\alpha_p$, $(\reff/\rfree)^2\approx 4\alpha_p/3$.
The free wind's volume is a fraction $(\reff/\rfree)^{-3}$ of the whole 
wind bubble.  

Finally, under the assumption of spherical symmetry, the total 
energy in the bubble interior is
\begin{equation*}
    \Ebub = 4\pi \int_0^{\reff} 
    \left( \frac{3}{2}P + \frac{1}{2}\rho v^2  \right) r^2 d r \, .
\end{equation*}
Given that only the kinetic energy term contributes in the free wind, 
we can use \autoref{eq:free_wind} to derive the total energy in the 
free wind as $\pdot \rfree /2$. For the shocked wind we can 
derive the energy as
\begin{equation*}
     E_{\rm ps} = \frac{\pdot \rfree}{8} 
     \left[3 \left(\frac{\reff}{\rfree} \right)^3 
     - 2
     - \frac{\rfree}{\reff}  \right] \, .
\end{equation*}
Putting these together, the total energy in the bubble interior is 
\begin{equation}
    \label{eq:ebub_derive}
    \Ebub = \frac{\pdot\reff \ratr}{2} \, ,
\end{equation}
where 
\begin{equation}
    \label{eq:S_derive}
    \ratr = \frac{3}{4}\left(\frac{\reff}{\rfree} \right)^2 
    - \frac{1}{4}\left(\frac{\reff}{\rfree} \right)^{-2} + \frac{1}{2}\left(\frac{\reff}{\rfree}\right)^{-1}
\end{equation}
is the enhancement in energy above the case where the whole wind bubble 
is made up of free wind ($\reff=\rfree$ and $\ratr=1$). Using \autoref{eq:alphap_derive}, one can show that $\ratr \approx \alpha_p$ 
to within 6\%.

We note that to account for the shock between the free and post-shock wind 
being oblique\footnote{This is always the case, see Paper II and \autoref{fig:schematics} 
panel b.}, \autoref{eq:vps} can be multiplied  by a factor $4-3\mu^2$ to obtain the 
post-shock radial velocity, while for the post-shock pressure \autoref{eq:Pps} is 
multiplied by $\mu^2$. Here, $\mu$ is the average value of the dot product of the unit 
radial vector ($\hat{\mathbf{r}}$) with the unit normal to the shock 
($\hat{\mathbf{n}}_{\rm shock}$).


\bibliography{bibliography}{}

\begin{thebibliography}{}
\expandafter\ifx\csname natexlab\endcsname\relax\def\natexlab#1{#1}\fi
\providecommand{\url}[1]{\href{#1}{#1}}
\providecommand{\dodoi}[1]{doi:~\href{http://doi.org/#1}{\nolinkurl{#1}}}
\providecommand{\doeprint}[1]{\href{http://ascl.net/#1}{\nolinkurl{http://ascl.net/#1}}}
\providecommand{\doarXiv}[1]{\href{https://arxiv.org/abs/#1}{\nolinkurl{https://arxiv.org/abs/#1}}}

\bibitem[{{Avedisova}(1972)}]{Avedisova72}
{Avedisova}, V.~S. 1972, \sovast, 15, 708

\bibitem[{{Badjin} {et~al.}(2016){Badjin}, {Glazyrin}, {Manukovskiy}, \&
  {Blinnikov}}]{Badjin16}
{Badjin}, D.~A., {Glazyrin}, S.~I., {Manukovskiy}, K.~V., \& {Blinnikov}, S.~I.
  2016, \mnras, 459, 2188, \dodoi{10.1093/mnras/stw790}

\bibitem[{{Bally}(2016)}]{Bally16}
{Bally}, J. 2016, \araa, 54, 491, \dodoi{10.1146/annurev-astro-081915-023341}

\bibitem[{{Barnes} {et~al.}(2017){Barnes}, {Longmore}, {Battersby}, {Bally},
  {Kruijssen}, {Henshaw}, \& {Walker}}]{Barnes17}
{Barnes}, A.~T., {Longmore}, S.~N., {Battersby}, C., {et~al.} 2017, \mnras,
  469, 2263, \dodoi{10.1093/mnras/stx941}

\bibitem[{{Barnes} {et~al.}(2020){Barnes}, {Longmore}, {Dale}, {Krumholz},
  {Kruijssen}, \& {Bigiel}}]{Barnes20}
{Barnes}, A.~T., {Longmore}, S.~N., {Dale}, J.~E., {et~al.} 2020, \mnras, 498,
  4906, \dodoi{10.1093/mnras/staa2719}

\bibitem[{{Bastian} \& {Lardo}(2018)}]{BastianLardo18}
{Bastian}, N., \& {Lardo}, C. 2018, \araa, 56, 83,
  \dodoi{10.1146/annurev-astro-081817-051839}

\bibitem[{{Blondin} {et~al.}(1998){Blondin}, {Wright}, {Borkowski}, \&
  {Reynolds}}]{Blondin98}
{Blondin}, J.~M., {Wright}, E.~B., {Borkowski}, K.~J., \& {Reynolds}, S.~P.
  1998, \apj, 500, 342, \dodoi{10.1086/305708}

\bibitem[{{Bucciantini} {et~al.}(2004){Bucciantini}, {Amato}, {Bandiera},
  {Blondin}, \& {Del Zanna}}]{Bucciantini04}
{Bucciantini}, N., {Amato}, E., {Bandiera}, R., {Blondin}, J.~M., \& {Del
  Zanna}, L. 2004, \aap, 423, 253, \dodoi{10.1051/0004-6361:20040360}

\bibitem[{{Capriotti} \& {Kozminski}(2001)}]{CapriottiKozminski01}
{Capriotti}, E.~R., \& {Kozminski}, J.~F. 2001, \pasp, 113, 677,
  \dodoi{10.1086/320809}

\bibitem[{{Castor} {et~al.}(1975){Castor}, {McCray}, \& {Weaver}}]{Castor75}
{Castor}, J., {McCray}, R., \& {Weaver}, R. 1975, \apjl, 200, L107,
  \dodoi{10.1086/181908}

\bibitem[{{Chevance} {et~al.}(2020{\natexlab{a}}){Chevance}, {Kruijssen},
  {Vazquez-Semadeni}, {Nakamura}, {Klessen}, {Ballesteros-Paredes}, {Inutsuka},
  {Adamo}, \& {Hennebelle}}]{chevance20a}
{Chevance}, M., {Kruijssen}, J.~M.~D., {Vazquez-Semadeni}, E., {et~al.}
  2020{\natexlab{a}}, \ssr, 216, 50, \dodoi{10.1007/s11214-020-00674-x}

\bibitem[{{Chevance} {et~al.}(2020{\natexlab{b}}){Chevance}, {Kruijssen},
  {Krumholz}, {Groves}, {Keller}, {Hughes}, {Glover}, {Henshaw}, {Herrera},
  {Kim}, {Leroy}, {Pety}, {Razza}, {Rosolowsky}, {Schinnerer}, {Schruba},
  {Barnes}, {Bigiel}, {Blanc}, {Emsellem}, {Faesi}, {Grasha}, {Klessen},
  {Kreckel}, {Liu}, {Longmore}, {Meidt}, {Querejeta}, {Saito}, {Sun}, \&
  {Usero}}]{chevance20}
{Chevance}, M., {Kruijssen}, J.~M.~D., {Krumholz}, M.~R., {et~al.}
  2020{\natexlab{b}}, arXiv e-prints, arXiv:2010.13788.
\newblock \doarXiv{2010.13788}

\bibitem[{{Cowie} \& {McKee}(1977)}]{CowieMcKee77}
{Cowie}, L.~L., \& {McKee}, C.~F. 1977, \apj, 211, 135, \dodoi{10.1086/154911}

\bibitem[{{Dale}(2017)}]{dale17}
{Dale}, J.~E. 2017, \mnras, 467, 1067, \dodoi{10.1093/mnras/stx028}

\bibitem[{{Dale} {et~al.}(2012){Dale}, {Ercolano}, \& {Bonnell}}]{Dale12}
{Dale}, J.~E., {Ercolano}, B., \& {Bonnell}, I.~A. 2012, \mnras, 424, 377,
  \dodoi{10.1111/j.1365-2966.2012.21205.x}

\bibitem[{{Dale} {et~al.}(2014){Dale}, {Ngoumou}, {Ercolano}, \&
  {Bonnell}}]{Dale14}
{Dale}, J.~E., {Ngoumou}, J., {Ercolano}, B., \& {Bonnell}, I.~A. 2014, \mnras,
  442, 694, \dodoi{10.1093/mnras/stu816}

\bibitem[{{Dunne} {et~al.}(2003){Dunne}, {Chu}, {Chen}, {Lowry}, {Townsley},
  {Gruendl}, {Guerrero}, \& {Rosado}}]{Dunne03}
{Dunne}, B.~C., {Chu}, Y.-H., {Chen}, C. H.~R., {et~al.} 2003, \apj, 590, 306,
  \dodoi{10.1086/375010}

\bibitem[{{El-Badry} {et~al.}(2019){El-Badry}, {Ostriker}, {Kim}, {Quataert},
  \& {Weisz}}]{ElBadry19}
{El-Badry}, K., {Ostriker}, E.~C., {Kim}, C.-G., {Quataert}, E., \& {Weisz},
  D.~R. 2019, \mnras, 490, 1961, \dodoi{10.1093/mnras/stz2773}

\bibitem[{{Emig} {et~al.}(2020){Emig}, {Bolatto}, {Leroy}, {Mills},
  {Jim{\'e}nez Donaire}, {Tielens}, {Ginsburg}, {Gorski}, {Krieger}, {Levy},
  {Meier}, {Ott}, {Rosolowsky}, {Thompson}, \& {Veilleux}}]{Emig20}
{Emig}, K.~L., {Bolatto}, A.~D., {Leroy}, A.~K., {et~al.} 2020, \apj, 903, 50,
  \dodoi{10.3847/1538-4357/abb67d}

\bibitem[{{Evans} {et~al.}(2009){Evans}, {Dunham}, {J{\o}rgensen}, {Enoch},
  {Mer{\'\i}n}, {van Dishoeck}, {Alcal{\'a}}, {Myers}, {Stapelfeldt}, {Huard},
  {Allen}, {Harvey}, {van Kempen}, {Blake}, {Koerner}, {Mundy}, {Padgett}, \&
  {Sargent}}]{Evans09}
{Evans}, Neal~J., I., {Dunham}, M.~M., {J{\o}rgensen}, J.~K., {et~al.} 2009,
  \apjs, 181, 321, \dodoi{10.1088/0067-0049/181/2/321}

\bibitem[{{Fall} {et~al.}(2010){Fall}, {Krumholz}, \& {Matzner}}]{Fall10}
{Fall}, S.~M., {Krumholz}, M.~R., \& {Matzner}, C.~D. 2010, \apjl, 710, L142,
  \dodoi{10.1088/2041-8205/710/2/L142}

\bibitem[{{Fielding} {et~al.}(2020){Fielding}, {Ostriker}, {Bryan}, \&
  {Jermyn}}]{FieldingFractal20}
{Fielding}, D.~B., {Ostriker}, E.~C., {Bryan}, G.~L., \& {Jermyn}, A.~S. 2020,
  \apjl, 894, L24, \dodoi{10.3847/2041-8213/ab8d2c}

\bibitem[{{Folini} \& {Walder}(2006)}]{FoliniWalder06}
{Folini}, D., \& {Walder}, R. 2006, \aap, 459, 1,
  \dodoi{10.1051/0004-6361:20053898}

\bibitem[{{Franco} {et~al.}(1994){Franco}, {Shore}, \&
  {Tenorio-Tagle}}]{franco94}
{Franco}, J., {Shore}, S.~N., \& {Tenorio-Tagle}, G. 1994, \apj, 436, 795,
  \dodoi{10.1086/174955}

\bibitem[{{Fukushima} {et~al.}(2020){Fukushima}, {Yajima}, {Sugimura},
  {Hosokawa}, {Omukai}, \& {Matsumoto}}]{Fukushima20}
{Fukushima}, H., {Yajima}, H., {Sugimura}, K., {et~al.} 2020, \mnras, 497,
  3830, \dodoi{10.1093/mnras/staa2062}

\bibitem[{{Gagn{\'e}} {et~al.}(2005){Gagn{\'e}}, {Oksala}, {Cohen}, {Tonnesen},
  {ud-Doula}, {Owocki}, {Townsend}, \& {MacFarlane}}]{Gagne05}
{Gagn{\'e}}, M., {Oksala}, M.~E., {Cohen}, D.~H., {et~al.} 2005, \apj, 628,
  986, \dodoi{10.1086/430873}

\bibitem[{{Garcia-Segura} {et~al.}(1996){Garcia-Segura}, {Langer}, \& {Mac
  Low}}]{GarciaSegura96}
{Garcia-Segura}, G., {Langer}, N., \& {Mac Low}, M.~M. 1996, \aap, 316, 133

\bibitem[{{Gatto} {et~al.}(2017){Gatto}, {Walch}, {Naab}, {Girichidis},
  {W{\"u}nsch}, {Glover}, {Klessen}, {Clark}, {Peters}, {Derigs}, {Baczynski},
  \& {Puls}}]{Gatto17}
{Gatto}, A., {Walch}, S., {Naab}, T., {et~al.} 2017, \mnras, 466, 1903,
  \dodoi{10.1093/mnras/stw3209}

\bibitem[{{Geen} {et~al.}(2020){Geen}, {Bieri}, {Rosdahl}, \& {de
  Koter}}]{Geen20}
{Geen}, S., {Bieri}, R., {Rosdahl}, J., \& {de Koter}, A. 2020, arXiv e-prints,
  arXiv:2009.08742.
\newblock \doarXiv{2009.08742}

\bibitem[{{Geen} {et~al.}(2015{\natexlab{a}}){Geen}, {Hennebelle}, {Tremblin},
  \& {Rosdahl}}]{Geen15b}
{Geen}, S., {Hennebelle}, P., {Tremblin}, P., \& {Rosdahl}, J.
  2015{\natexlab{a}}, \mnras, 454, 4484, \dodoi{10.1093/mnras/stv2272}

\bibitem[{{Geen} {et~al.}(2015{\natexlab{b}}){Geen}, {Rosdahl}, {Blaizot},
  {Devriendt}, \& {Slyz}}]{Geen15a}
{Geen}, S., {Rosdahl}, J., {Blaizot}, J., {Devriendt}, J., \& {Slyz}, A.
  2015{\natexlab{b}}, \mnras, 448, 3248, \dodoi{10.1093/mnras/stv251}

\bibitem[{{Geen} {et~al.}(2017){Geen}, {Soler}, \& {Hennebelle}}]{Geen17}
{Geen}, S., {Soler}, J.~D., \& {Hennebelle}, P. 2017, \mnras, 471, 4844,
  \dodoi{10.1093/mnras/stx1765}

\bibitem[{{Girichidis} {et~al.}(2020){Girichidis}, {Offner}, {Kritsuk},
  {Klessen}, {Hennebelle}, {Kruijssen}, {Krause}, {Glover}, \&
  {Padovani}}]{girichidis20}
{Girichidis}, P., {Offner}, S. S.~R., {Kritsuk}, A.~G., {et~al.} 2020, \ssr,
  216, 68, \dodoi{10.1007/s11214-020-00693-8}

\bibitem[{{Gratton} {et~al.}(2019){Gratton}, {Bragaglia}, {Carretta},
  {D'Orazi}, {Lucatello}, \& {Sollima}}]{Gratton19}
{Gratton}, R., {Bragaglia}, A., {Carretta}, E., {et~al.} 2019, \aapr, 27, 8,
  \dodoi{10.1007/s00159-019-0119-3}

\bibitem[{{Gronke} \& {Oh}(2018{\natexlab{a}})}]{GronkeOh18}
{Gronke}, M., \& {Oh}, S.~P. 2018{\natexlab{a}}, \mnras, 480, L111,
  \dodoi{10.1093/mnrasl/sly131}

\bibitem[{{Gronke} \& {Oh}(2018{\natexlab{b}})}]{Gronke2018}
---. 2018{\natexlab{b}}, \mnras, 480, L111, \dodoi{10.1093/mnrasl/sly131}

\bibitem[{{Grudi{\'c}} {et~al.}(2018){Grudi{\'c}}, {Hopkins},
  {Faucher-Gigu{\`e}re}, {Quataert}, {Murray}, \& {Kere{\v{s}}}}]{Grudic18}
{Grudi{\'c}}, M.~Y., {Hopkins}, P.~F., {Faucher-Gigu{\`e}re}, C.-A., {et~al.}
  2018, \mnras, 475, 3511, \dodoi{10.1093/mnras/sty035}

\bibitem[{{G{\"u}del} {et~al.}(2008){G{\"u}del}, {Briggs}, {Montmerle},
  {Audard}, {Rebull}, \& {Skinner}}]{Gudel08}
{G{\"u}del}, M., {Briggs}, K.~R., {Montmerle}, T., {et~al.} 2008, Science, 319,
  309, \dodoi{10.1126/science.1149926}

\bibitem[{{Haid} {et~al.}(2018){Haid}, {Walch}, {Seifried}, {W{\"u}nsch},
  {Dinnbier}, \& {Naab}}]{Haid18}
{Haid}, S., {Walch}, S., {Seifried}, D., {et~al.} 2018, \mnras, 478, 4799,
  \dodoi{10.1093/mnras/sty1315}

\bibitem[{{Harper-Clark} \& {Murray}(2009)}]{HCM09}
{Harper-Clark}, E., \& {Murray}, N. 2009, \apj, 693, 1696,
  \dodoi{10.1088/0004-637X/693/2/1696}

\bibitem[{{Harris} {et~al.}(2020){Harris}, {Jarrod Millman}, {van der Walt},
  {Gommers}, {Virtanen}, {Cournapeau}, {Wieser}, {Taylor}, {Berg}, {Smith},
  {Kern}, {Picus}, {Hoyer}, {van Kerkwijk}, {Brett}, {Haldane}, {Fern{\'a}ndez
  del R{\'\i}o}, {Wiebe}, {Peterson}, {G{\'e}rard-Marchant}, {Sheppard},
  {Reddy}, {Weckesser}, {Abbasi}, {Gohlke}, \& {Oliphant}}]{harrisNumpy2020}
{Harris}, C.~R., {Jarrod Millman}, K., {van der Walt}, S.~J., {et~al.} 2020,
  arXiv e-prints, arXiv:2006.10256.
\newblock \doarXiv{2006.10256}

\bibitem[{{He} {et~al.}(2019){He}, {Ricotti}, \& {Geen}}]{he19}
{He}, C.-C., {Ricotti}, M., \& {Geen}, S. 2019, \mnras, 489, 1880,
  \dodoi{10.1093/mnras/stz2239}

\bibitem[{{Hennebelle} \& {Iffrig}(2014)}]{hennebelle14}
{Hennebelle}, P., \& {Iffrig}, O. 2014, \aap, 570, A81,
  \dodoi{10.1051/0004-6361/201423392}

\bibitem[{{Hosek} {et~al.}(2019){Hosek}, {Lu}, {Anderson}, {Najarro}, {Ghez},
  {Morris}, {Clarkson}, \& {Albers}}]{Hosek19}
{Hosek}, Matthew~W., J., {Lu}, J.~R., {Anderson}, J., {et~al.} 2019, \apj, 870,
  44, \dodoi{10.3847/1538-4357/aaef90}

\bibitem[{{Hunter}(2007)}]{matplotlib_hunter07}
{Hunter}, J.~D. 2007, Computing in Science and Engineering, 9, 90,
  \dodoi{10.1109/MCSE.2007.55}

\bibitem[{{Johnson} {et~al.}(2015){Johnson}, {Leroy}, {Indebetouw}, {Brogan},
  {Whitmore}, {Hibbard}, {Sheth}, \& {Evans}}]{Johnson15}
{Johnson}, K.~E., {Leroy}, A.~K., {Indebetouw}, R., {et~al.} 2015, \apj, 806,
  35, \dodoi{10.1088/0004-637X/806/1/35}

\bibitem[{{Kennicutt}(1998)}]{KennicuttRev98}
{Kennicutt}, Robert~C., J. 1998, \araa, 36, 189,
  \dodoi{10.1146/annurev.astro.36.1.189}

\bibitem[{{Kim} \& {Ostriker}(2017)}]{CGK_TIGRESS1}
{Kim}, C.-G., \& {Ostriker}, E.~C. 2017, \apj, 846, 133,
  \dodoi{10.3847/1538-4357/aa8599}

\bibitem[{{Kim} {et~al.}(2013){Kim}, {Ostriker}, \& {Kim}}]{kok13}
{Kim}, C.-G., {Ostriker}, E.~C., \& {Kim}, W.-T. 2013, \apj, 776, 1,
  \dodoi{10.1088/0004-637X/776/1/1}

\bibitem[{{Kim} {et~al.}(2017){Kim}, {Ostriker}, \&
  {Raileanu}}]{KimOstrikerRaileanu17}
{Kim}, C.-G., {Ostriker}, E.~C., \& {Raileanu}, R. 2017, \apj, 834, 25,
  \dodoi{10.3847/1538-4357/834/1/25}

\bibitem[{{Kim} {et~al.}(2016){Kim}, {Kim}, \& {Ostriker}}]{JGK16}
{Kim}, J.-G., {Kim}, W.-T., \& {Ostriker}, E.~C. 2016, \apj, 819, 137,
  \dodoi{10.3847/0004-637X/819/2/137}

\bibitem[{{Kim} {et~al.}(2018){Kim}, {Kim}, \& {Ostriker}}]{JGK18}
---. 2018, \apj, 859, 68, \dodoi{10.3847/1538-4357/aabe27}

\bibitem[{{Kim} {et~al.}(2019){Kim}, {Kim}, \& {Ostriker}}]{kim19}
---. 2019, \apj, 883, 102, \dodoi{10.3847/1538-4357/ab3d3d}

\bibitem[{{Kim} {et~al.}(2020){Kim}, {Ostriker}, \& {Filippova}}]{JGK20}
{Kim}, J.-G., {Ostriker}, E.~C., \& {Filippova}, N. 2020, arXiv e-prints,
  arXiv:2011.07772.
\newblock \doarXiv{2011.07772}

\bibitem[{{Klein} {et~al.}(1994){Klein}, {McKee}, \& {Colella}}]{Klein94}
{Klein}, R.~I., {McKee}, C.~F., \& {Colella}, P. 1994, \apj, 420, 213,
  \dodoi{10.1086/173554}

\bibitem[{{Koo} \& {McKee}(1992{\natexlab{a}})}]{KooMcKee92a}
{Koo}, B.-C., \& {McKee}, C.~F. 1992{\natexlab{a}}, \apj, 388, 93,
  \dodoi{10.1086/171132}

\bibitem[{{Koo} \& {McKee}(1992{\natexlab{b}})}]{KooMcKee92b}
---. 1992{\natexlab{b}}, \apj, 388, 103, \dodoi{10.1086/171133}

\bibitem[{{Kroupa}(2001)}]{KroupaIMF}
{Kroupa}, P. 2001, \mnras, 322, 231, \dodoi{10.1046/j.1365-8711.2001.04022.x}

\bibitem[{{Kruijssen} {et~al.}(2019){Kruijssen}, {Schruba}, {Chevance},
  {Longmore}, {Hygate}, {Haydon}, {McLeod}, {Dalcanton}, {Tacconi}, \& {van
  Dishoeck}}]{Kruijssen19}
{Kruijssen}, J.~M.~D., {Schruba}, A., {Chevance}, M., {et~al.} 2019, \nat, 569,
  519, \dodoi{10.1038/s41586-019-1194-3}

\bibitem[{{Krumholz} \& {Matzner}(2009)}]{KrumholzMatzner09}
{Krumholz}, M.~R., \& {Matzner}, C.~D. 2009, \apj, 703, 1352,
  \dodoi{10.1088/0004-637X/703/2/1352}

\bibitem[{{Krumholz} {et~al.}(2019){Krumholz}, {McKee}, \& {Bland
  -Hawthorn}}]{KMBBH19}
{Krumholz}, M.~R., {McKee}, C.~F., \& {Bland -Hawthorn}, J. 2019, \araa, 57,
  227, \dodoi{10.1146/annurev-astro-091918-104430}

\bibitem[{{Krumholz} \& {Tan}(2007)}]{KrumholzTan07}
{Krumholz}, M.~R., \& {Tan}, J.~C. 2007, \apj, 654, 304, \dodoi{10.1086/509101}

\bibitem[{{Lada} {et~al.}(2010){Lada}, {Lombardi}, \& {Alves}}]{Lada2010}
{Lada}, C.~J., {Lombardi}, M., \& {Alves}, J.~F. 2010, \apj, 724, 687,
  \dodoi{10.1088/0004-637X/724/1/687}

\bibitem[{{Lancaster} {et~al.}(2021){Lancaster}, {Ostriker}, {Kim}, \&
  {Kim}}]{PaperII}
{Lancaster}, L., {Ostriker}, E.~C., {Kim}, J.-G., \& {Kim}, C.-G. 2021, \apj

\bibitem[{{Lee} {et~al.}(2016){Lee}, {Miville-Desch{\^e}nes}, \&
  {Murray}}]{Lee16}
{Lee}, E.~J., {Miville-Desch{\^e}nes}, M.-A., \& {Murray}, N.~W. 2016, \apj,
  833, 229, \dodoi{10.3847/1538-4357/833/2/229}

\bibitem[{{Leitherer} {et~al.}(2014){Leitherer}, {Ekstr{\"o}m}, {Meynet},
  {Schaerer}, {Agienko}, \& {Levesque}}]{Leitherer14}
{Leitherer}, C., {Ekstr{\"o}m}, S., {Meynet}, G., {et~al.} 2014, \apjs, 212,
  14, \dodoi{10.1088/0067-0049/212/1/14}

\bibitem[{{Leitherer} {et~al.}(1992){Leitherer}, {Robert}, \&
  {Drissen}}]{Leitherer92}
{Leitherer}, C., {Robert}, C., \& {Drissen}, L. 1992, \apj, 401, 596,
  \dodoi{10.1086/172089}

\bibitem[{{Leitherer} {et~al.}(1999){Leitherer}, {Schaerer}, {Goldader},
  {Delgado}, {Robert}, {Kune}, {de Mello}, {Devost}, \& {Heckman}}]{SB99}
{Leitherer}, C., {Schaerer}, D., {Goldader}, J.~D., {et~al.} 1999, \apjs, 123,
  3, \dodoi{10.1086/313233}

\bibitem[{{Leroy} {et~al.}(2017){Leroy}, {Schinnerer}, {Hughes}, {Kruijssen},
  {Meidt}, {Schruba}, {Sun}, {Bigiel}, {Aniano}, {Blanc}, {Bolatto},
  {Chevance}, {Colombo}, {Gallagher}, {Garcia-Burillo}, {Kramer}, {Querejeta},
  {Pety}, {Thompson}, \& {Usero}}]{Leroy17}
{Leroy}, A.~K., {Schinnerer}, E., {Hughes}, A., {et~al.} 2017, \apj, 846, 71,
  \dodoi{10.3847/1538-4357/aa7fef}

\bibitem[{{Leroy} {et~al.}(2018){Leroy}, {Bolatto}, {Ostriker}, {Walter},
  {Gorski}, {Ginsburg}, {Krieger}, {Levy}, {Meier}, {Mills}, {Ott},
  {Rosolowsky}, {Thompson}, {Veilleux}, \& {Zschaechner}}]{Leroy18}
{Leroy}, A.~K., {Bolatto}, A.~D., {Ostriker}, E.~C., {et~al.} 2018, \apj, 869,
  126, \dodoi{10.3847/1538-4357/aaecd1}

\bibitem[{{Levy} {et~al.}(2020){Levy}, {Bolatto}, {Leroy}, {Emig}, {Gorski},
  {Krieger}, {Lenkic}, {Meier}, {Mills}, {Ott}, {Rosolowsky}, {Tarantino},
  {Veilleux}, {Walter}, {Weiss}, \& {Zwaan}}]{Levy20}
{Levy}, R.~C., {Bolatto}, A.~D., {Leroy}, A.~K., {et~al.} 2020, arXiv e-prints,
  arXiv:2011.05334.
\newblock \doarXiv{2011.05334}

\bibitem[{{Li} {et~al.}(2019){Li}, {Vogelsberger}, {Marinacci}, \&
  {Gnedin}}]{Li19}
{Li}, H., {Vogelsberger}, M., {Marinacci}, F., \& {Gnedin}, O.~Y. 2019, \mnras,
  487, 364, \dodoi{10.1093/mnras/stz1271}

\bibitem[{{Lochhaas} \& {Thompson}(2017)}]{LochhaasThompson17}
{Lochhaas}, C., \& {Thompson}, T.~A. 2017, \mnras, 470, 977,
  \dodoi{10.1093/mnras/stx1289}

\bibitem[{{Lopez} {et~al.}(2011){Lopez}, {Krumholz}, {Bolatto}, {Prochaska}, \&
  {Ramirez-Ruiz}}]{Lopez11}
{Lopez}, L.~A., {Krumholz}, M.~R., {Bolatto}, A.~D., {Prochaska}, J.~X., \&
  {Ramirez-Ruiz}, E. 2011, \apj, 731, 91, \dodoi{10.1088/0004-637X/731/2/91}

\bibitem[{{Lopez} {et~al.}(2014){Lopez}, {Krumholz}, {Bolatto}, {Prochaska},
  {Ramirez-Ruiz}, \& {Castro}}]{Lopez14}
{Lopez}, L.~A., {Krumholz}, M.~R., {Bolatto}, A.~D., {et~al.} 2014, \apj, 795,
  121, \dodoi{10.1088/0004-637X/795/2/121}

\bibitem[{{Lu} {et~al.}(2013){Lu}, {Do}, {Ghez}, {Morris}, {Yelda}, \&
  {Matthews}}]{Lu13}
{Lu}, J.~R., {Do}, T., {Ghez}, A.~M., {et~al.} 2013, \apj, 764, 155,
  \dodoi{10.1088/0004-637X/764/2/155}

\bibitem[{{Mac Low} \& {McCray}(1988)}]{maclow88}
{Mac Low}, M.-M., \& {McCray}, R. 1988, \apj, 324, 776, \dodoi{10.1086/165936}

\bibitem[{{Mackey} {et~al.}(2015){Mackey}, {Gvaramadze}, {Mohamed}, \&
  {Langer}}]{Mackey15}
{Mackey}, J., {Gvaramadze}, V.~V., {Mohamed}, S., \& {Langer}, N. 2015, \aap,
  573, A10, \dodoi{10.1051/0004-6361/201424716}

\bibitem[{{Matzner}(2002)}]{Matzner02}
{Matzner}, C.~D. 2002, \apj, 566, 302, \dodoi{10.1086/338030}

\bibitem[{{Matzner} \& {McKee}(2000)}]{MatznerMcKee00}
{Matzner}, C.~D., \& {McKee}, C.~F. 2000, \apj, 545, 364,
  \dodoi{10.1086/317785}

\bibitem[{{McCrady} {et~al.}(2005){McCrady}, {Graham}, \& {Vacca}}]{McCrady05}
{McCrady}, N., {Graham}, J.~R., \& {Vacca}, W.~D. 2005, \apj, 621, 278,
  \dodoi{10.1086/427487}

\bibitem[{{McCray} \& {Kafatos}(1987)}]{McCrayKafatos87}
{McCray}, R., \& {Kafatos}, M. 1987, \apj, 317, 190, \dodoi{10.1086/165267}

\bibitem[{{McLeod} {et~al.}(2019){McLeod}, {Dale}, {Evans}, {Ginsburg},
  {Kruijssen}, {Pellegrini}, {Ramsay}, \& {Testi}}]{McLeod19}
{McLeod}, A.~F., {Dale}, J.~E., {Evans}, C.~J., {et~al.} 2019, \mnras, 486,
  5263, \dodoi{10.1093/mnras/sty2696}

\bibitem[{{McLeod} {et~al.}(2020){McLeod}, {Kruijssen}, {Weisz}, {Zeidler},
  {Schruba}, {Dalcanton}, {Longmore}, {Chevance}, {Faesi}, \&
  {Byler}}]{McLeod20}
{McLeod}, A.~F., {Kruijssen}, J.~M.~D., {Weisz}, D.~R., {et~al.} 2020, \apj,
  891, 25, \dodoi{10.3847/1538-4357/ab6d63}

\bibitem[{{Michaut} {et~al.}(2012){Michaut}, {Cavet}, {Bouquet}, {Roy}, \&
  {Nguyen}}]{Michaut12}
{Michaut}, C., {Cavet}, C., {Bouquet}, S.~E., {Roy}, F., \& {Nguyen}, H.~C.
  2012, \apj, 759, 78, \dodoi{10.1088/0004-637X/759/2/78}

\bibitem[{{Murray}(2011)}]{Murray11}
{Murray}, N. 2011, \apj, 729, 133, \dodoi{10.1088/0004-637X/729/2/133}

\bibitem[{{Murray} {et~al.}(2010){Murray}, {Quataert}, \&
  {Thompson}}]{Murray10}
{Murray}, N., {Quataert}, E., \& {Thompson}, T.~A. 2010, \apj, 709, 191,
  \dodoi{10.1088/0004-637X/709/1/191}

\bibitem[{{Ntormousi} {et~al.}(2011){Ntormousi}, {Burkert}, {Fierlinger}, \&
  {Heitsch}}]{Ntormousi11}
{Ntormousi}, E., {Burkert}, A., {Fierlinger}, K., \& {Heitsch}, F. 2011, \apj,
  731, 13, \dodoi{10.1088/0004-637X/731/1/13}

\bibitem[{{Oey} {et~al.}(2017){Oey}, {Herrera}, {Silich}, {Reiter}, {James},
  {Jaskot}, \& {Micheva}}]{Oey17}
{Oey}, M.~S., {Herrera}, C.~N., {Silich}, S., {et~al.} 2017, \apjl, 849, L1,
  \dodoi{10.3847/2041-8213/aa9215}

\bibitem[{{Offner} \& {Chaban}(2017)}]{OffnerChaban17}
{Offner}, S. S.~R., \& {Chaban}, J. 2017, \apj, 847, 104,
  \dodoi{10.3847/1538-4357/aa8996}

\bibitem[{{Olivier} {et~al.}(2020){Olivier}, {Lopez}, {Rosen}, {Nayak},
  {Rieter}, {Krumholz}, \& {Bolatto}}]{Olivier20}
{Olivier}, G.~M., {Lopez}, L.~A., {Rosen}, A.~L., {et~al.} 2020, arXiv
  e-prints, arXiv:2009.10079.
\newblock \doarXiv{2009.10079}

\bibitem[{{Ostriker} \& {McKee}(1988)}]{OstrikerMcKee88}
{Ostriker}, J.~P., \& {McKee}, C.~F. 1988, Reviews of Modern Physics, 60, 1,
  \dodoi{10.1103/RevModPhys.60.1}

\bibitem[{{Pellegrini} {et~al.}(2011){Pellegrini}, {Baldwin}, \&
  {Ferland}}]{pellegrini11}
{Pellegrini}, E.~W., {Baldwin}, J.~A., \& {Ferland}, G.~J. 2011, \apj, 738, 34,
  \dodoi{10.1088/0004-637X/738/1/34}

\bibitem[{{Pittard}(2013)}]{Pittard13}
{Pittard}, J.~M. 2013, \mnras, 435, 3600, \dodoi{10.1093/mnras/stt1552}

\bibitem[{{Portegies Zwart} {et~al.}(2010){Portegies Zwart}, {McMillan}, \&
  {Gieles}}]{Portegies-Zwart10}
{Portegies Zwart}, S.~F., {McMillan}, S. L.~W., \& {Gieles}, M. 2010, \araa,
  48, 431, \dodoi{10.1146/annurev-astro-081309-130834}

\bibitem[{{Rahner} {et~al.}(2017){Rahner}, {Pellegrini}, {Glover}, \&
  {Klessen}}]{Rahner17}
{Rahner}, D., {Pellegrini}, E.~W., {Glover}, S. C.~O., \& {Klessen}, R.~S.
  2017, \mnras, 470, 4453, \dodoi{10.1093/mnras/stx1532}

\bibitem[{{Rahner} {et~al.}(2019){Rahner}, {Pellegrini}, {Glover}, \&
  {Klessen}}]{Rahner19}
---. 2019, \mnras, 483, 2547, \dodoi{10.1093/mnras/sty3295}

\bibitem[{{Raskutti} {et~al.}(2016){Raskutti}, {Ostriker}, \&
  {Skinner}}]{Raskutti16}
{Raskutti}, S., {Ostriker}, E.~C., \& {Skinner}, M.~A. 2016, \apj, 829, 130,
  \dodoi{10.3847/0004-637X/829/2/130}

\bibitem[{{Raskutti} {et~al.}(2017){Raskutti}, {Ostriker}, \&
  {Skinner}}]{raskutti17}
---. 2017, \apj, 850, 112, \dodoi{10.3847/1538-4357/aa965e}

\bibitem[{{Rogers} \& {Pittard}(2013)}]{RogersPittard13}
{Rogers}, H., \& {Pittard}, J.~M. 2013, \mnras, 431, 1337,
  \dodoi{10.1093/mnras/stt255}

\bibitem[{{Rosen} {et~al.}(2014){Rosen}, {Lopez}, {Krumholz}, \&
  {Ramirez-Ruiz}}]{Rosen14}
{Rosen}, A.~L., {Lopez}, L.~A., {Krumholz}, M.~R., \& {Ramirez-Ruiz}, E. 2014,
  \mnras, 442, 2701, \dodoi{10.1093/mnras/stu1037}

\bibitem[{{Sano} {et~al.}(2012){Sano}, {Nishihara}, {Matsuoka}, \&
  {Inoue}}]{Sano12}
{Sano}, T., {Nishihara}, K., {Matsuoka}, C., \& {Inoue}, T. 2012, \apj, 758,
  126, \dodoi{10.1088/0004-637X/758/2/126}

\bibitem[{{Scannapieco} \& {Br{\"u}ggen}(2015)}]{Scannapieco2015}
{Scannapieco}, E., \& {Br{\"u}ggen}, M. 2015, \apj, 805, 158,
  \dodoi{10.1088/0004-637X/805/2/158}

\bibitem[{{Schneider} \& {Robertson}(2017)}]{Schneider2017}
{Schneider}, E.~E., \& {Robertson}, B.~E. 2017, \apj, 834, 144,
  \dodoi{10.3847/1538-4357/834/2/144}

\bibitem[{{Silich} \& {Tenorio-Tagle}(2013)}]{SilichTT13}
{Silich}, S., \& {Tenorio-Tagle}, G. 2013, \apj, 765, 43,
  \dodoi{10.1088/0004-637X/765/1/43}

\bibitem[{{Skinner} \& {Ostriker}(2015)}]{Skinner15}
{Skinner}, M.~A., \& {Ostriker}, E.~C. 2015, \apj, 809, 187,
  \dodoi{10.1088/0004-637X/809/2/187}

\bibitem[{{Steigman} {et~al.}(1975){Steigman}, {Strittmatter}, \&
  {Williams}}]{Steigman75}
{Steigman}, G., {Strittmatter}, P.~A., \& {Williams}, R.~E. 1975, \apj, 198,
  575, \dodoi{10.1086/153636}

\bibitem[{{Tan} {et~al.}(2020){Tan}, {Oh}, \& {Gronke}}]{Tan20}
{Tan}, B., {Oh}, S.~P., \& {Gronke}, M. 2020, arXiv e-prints, arXiv:2008.12302.
\newblock \doarXiv{2008.12302}

\bibitem[{{Thompson} {et~al.}(2005){Thompson}, {Quataert}, \&
  {Murray}}]{Thompson05}
{Thompson}, T.~A., {Quataert}, E., \& {Murray}, N. 2005, \apj, 630, 167,
  \dodoi{10.1086/431923}

\bibitem[{{Toal{\'a}} \& {Arthur}(2018)}]{ToalaArthur18}
{Toal{\'a}}, J.~A., \& {Arthur}, S.~J. 2018, \mnras, 478, 1218,
  \dodoi{10.1093/mnras/sty1127}

\bibitem[{{Townsley} {et~al.}(2011){Townsley}, {Broos}, {Chu}, {Gruendl},
  {Oey}, \& {Pittard}}]{Townsley11}
{Townsley}, L.~K., {Broos}, P.~S., {Chu}, Y.-H., {et~al.} 2011, \apjs, 194, 16,
  \dodoi{10.1088/0067-0049/194/1/16}

\bibitem[{{Townsley} {et~al.}(2006){Townsley}, {Broos}, {Feigelson}, {Brandl},
  {Chu}, {Garmire}, \& {Pavlov}}]{Townsley06}
{Townsley}, L.~K., {Broos}, P.~S., {Feigelson}, E.~D., {et~al.} 2006, \aj, 131,
  2140, \dodoi{10.1086/500532}

\bibitem[{{Townsley} {et~al.}(2003){Townsley}, {Feigelson}, {Montmerle},
  {Broos}, {Chu}, \& {Garmire}}]{Townsley03}
{Townsley}, L.~K., {Feigelson}, E.~D., {Montmerle}, T., {et~al.} 2003, \apj,
  593, 874, \dodoi{10.1086/376692}

\bibitem[{{Tsang} \& {Milosavljevi{\'c}}(2018)}]{Tsang18}
{Tsang}, B. T.~H., \& {Milosavljevi{\'c}}, M. 2018, \mnras, 478, 4142,
  \dodoi{10.1093/mnras/sty1217}

\bibitem[{{Turner} {et~al.}(2017){Turner}, {Consiglio}, {Beck}, {Goss}, {Ho},
  {Meier}, {Silich}, \& {Zhao}}]{Turner17}
{Turner}, J.~L., {Consiglio}, S.~M., {Beck}, S.~C., {et~al.} 2017, \apj, 846,
  73, \dodoi{10.3847/1538-4357/aa8669}

\bibitem[{{Utomo} {et~al.}(2018){Utomo}, {Sun}, {Leroy}, {Kruijssen},
  {Schinnerer}, {Schruba}, {Bigiel}, {Blanc}, {Chevance}, {Emsellem},
  {Herrera}, {Hygate}, {Kreckel}, {Ostriker}, {Pety}, {Querejeta},
  {Rosolowsky}, {Sandstrom}, \& {Usero}}]{Utomo18}
{Utomo}, D., {Sun}, J., {Leroy}, A.~K., {et~al.} 2018, \apjl, 861, L18,
  \dodoi{10.3847/2041-8213/aacf8f}

\bibitem[{{Vink} {et~al.}(2001){Vink}, {de Koter}, \& {Lamers}}]{Vink01}
{Vink}, J.~S., {de Koter}, A., \& {Lamers}, H.~J.~G.~L.~M. 2001, \aap, 369,
  574, \dodoi{10.1051/0004-6361:20010127}

\bibitem[{{Virtanen} {et~al.}(2020){Virtanen}, {Gommers}, {Oliphant},
  {Haberland}, {Reddy}, {Cournapeau}, {Burovski}, {Peterson}, {Weckesser},
  {Bright}, {van der Walt}, {Brett}, {Wilson}, {Millman}, {Mayorov}, {Nelson},
  {Jones}, {Kern}, {Larson}, {Carey}, {Polat}, {Feng}, {Moore}, {Vand erPlas},
  {Laxalde}, {Perktold}, {Cimrman}, {Henriksen}, {Quintero}, {Harris},
  {Archibald}, {Ribeiro}, {Pedregosa}, {van Mulbregt}, \& {SciPy 1. 0
  Contributors}}]{scipy}
{Virtanen}, P., {Gommers}, R., {Oliphant}, T.~E., {et~al.} 2020, Nature
  Methods, 17, 261, \dodoi{10.1038/s41592-019-0686-2}

\bibitem[{{Vishniac}(1983)}]{Vishniac83}
{Vishniac}, E.~T. 1983, \apj, 274, 152, \dodoi{10.1086/161433}

\bibitem[{{Vishniac}(1994)}]{Vishniac94}
---. 1994, \apj, 428, 186, \dodoi{10.1086/174231}

\bibitem[{{Vishniac} \& {Ryu}(1989)}]{VishniacRyu89}
{Vishniac}, E.~T., \& {Ryu}, D. 1989, \apj, 337, 917, \dodoi{10.1086/167161}

\bibitem[{{Vutisalchavakul} {et~al.}(2016){Vutisalchavakul}, {Evans}, \&
  {Heyer}}]{Vutisalchavakul16}
{Vutisalchavakul}, N., {Evans}, Neal~J., I., \& {Heyer}, M. 2016, \apj, 831,
  73, \dodoi{10.3847/0004-637X/831/1/73}

\bibitem[{{Weaver} {et~al.}(1977){Weaver}, {McCray}, {Castor}, {Shapiro}, \&
  {Moore}}]{Weaver77}
{Weaver}, R., {McCray}, R., {Castor}, J., {Shapiro}, P., \& {Moore}, R. 1977,
  \apj, 218, 377, \dodoi{10.1086/155692}

\bibitem[{{Whitmore}(2003)}]{Whitmore03}
{Whitmore}, B.~C. 2003, in A Decade of Hubble Space Telescope Science, ed.
  M.~{Livio}, K.~{Noll}, \& M.~{Stiavelli}, Vol.~14, 153--178

\bibitem[{{Whitworth}(1979)}]{Whitworth79}
{Whitworth}, A. 1979, \mnras, 186, 59, \dodoi{10.1093/mnras/186.1.59}

\bibitem[{{Wolfire} \& {Cassinelli}(1987)}]{wolfire87}
{Wolfire}, M.~G., \& {Cassinelli}, J.~P. 1987, \apj, 319, 850,
  \dodoi{10.1086/165503}

\bibitem[{{W{\"u}nsch} {et~al.}(2017){W{\"u}nsch}, {Palou{\v{s}}},
  {Tenorio-Tagle}, \& {Ehlerov{\'a}}}]{Wunsch17}
{W{\"u}nsch}, R., {Palou{\v{s}}}, J., {Tenorio-Tagle}, G., \& {Ehlerov{\'a}},
  S. 2017, \apj, 835, 60, \dodoi{10.3847/1538-4357/835/1/60}

\end{thebibliography}
\bibliographystyle{aasjournal}

\end{document}